\pdfoutput=1
\documentclass[11pt,a4paper]{scrartcl}

\usepackage[utf8]{inputenc}
\usepackage[bf]{caption}
\usepackage[T1]{fontenc} 
\usepackage[UKenglish]{babel} 
\usepackage[section]{placeins} 
\usepackage{todonotes}
\usepackage{a4, 
            amsfonts, 
            amsmath, 
            amssymb, 
            amsthm, 
            arydshln, 
            booktabs, 
            calligra, 
            colortbl, 
            etex, 
            fancyvrb, 
            hyperref, 
            lastpage, 
            listings, 
            longtable, 
            lscape, 
            mathrsfs, 
            pbox, 
            scrhack, 
            scrlayer-scrpage, 
            subcaption, 
            tikz, 
            tikz-cd, 
            xy 
}

\usetikzlibrary{matrix,arrows,automata,calc,chains,intersections,decorations.markings,matrix,shapes.geometric,positioning,plotmarks,scopes,patterns}
\xyoption{all}

\def\hybrid{\topmargin -20pt    \oddsidemargin 0pt
        \headheight 15.2pt \headsep 0pt
        \textwidth 6.25in       
        \textheight 9 in       
        \marginparwidth .875in
        \parskip 5pt plus 1pt 
        \jot = 1.5ex
   }
\hybrid

\numberwithin{equation}{section}
\numberwithin{table}{section}

\hypersetup{
    pdftitle={A Note on Non-Flat Points in the SU(5)xU(1)-PQ F-Theory Model},
    pdfauthor={Ismail Achmed-Zade, Inaki Garcia-Etxebarria, Christoph Mayrhofer},
    pdfsubject={Mordell-Weil groups in F-Theory, GUT models, non-flat fibrations}
}

\newcommand{\CP}{\mathbb{P}}
\newcommand{\Pnf}[1]{\CP^1_{\text{nf}_#1}}
\newcommand{\FS}{\mathcal{FS}}
\newcommand{\ten}{\mathbf{10}}
\newcommand{\five}{\mathbf{5}}
\newcommand{\singlet}{\mathbf{1}}

\newcommand{\bC}{\mathbb{C}}
\newcommand{\bP}{\mathbb{P}}
\newcommand{\Sp}{\ensuremath{U\!Sp}}

\begin{document}

\baselineskip=14pt
\parskip 5pt plus 1pt

\vspace*{-1.5cm}
\begin{flushright}    
  {\small
    MPP-2018-129 \\
    LMU-ASC 36/18
  }
\end{flushright}

\vspace{2cm}
\begin{center}        
  \textbf{{\LARGE A Note on Non-Flat Points\\[3mm] in the \texorpdfstring{$SU(5)\times U(1)_{PQ}$}{SU(5)xU(1)-PQ} F-Theory Model}}
\end{center}

\vspace{0.75cm}
\begin{center}        
Ismail Achmed-Zade\textsuperscript{1,2},  I\~naki Garc\'ia-Etxebarria\textsuperscript{3}, and Christoph Mayrhofer\textsuperscript{1}
\end{center}

\vspace{0.15cm}
\begin{center}        
  \emph{\textsuperscript{1} Arnold Sommerfeld Center for Theoretical Physics,\\
             Theresienstrasse 37, 80333 M\"unchen, Germany}
            \\[0.15cm]
  \emph{\textsuperscript{2} Max-Planck-Institut für Physik,\\
    Föhringer Ring 6, 80805 M\"unchen, Germany}\\[0.15cm]
  \emph{\textsuperscript{3} Department of Mathematical Sciences, Durham University,\\
    Durham, DH1 3LE, United Kingdom
  }
 \end{center}

\vspace{2cm}

\begin{abstract}
  Non-flat fibrations often appear in F-theory GUT models, and their
  interpretation is still somewhat mysterious. In this note we explore
  this issue in a model of particular phenomenological interest, the
  global $SU(5)\times U(1)$ Peccei-Quinn F-theory model. We present
  evidence that \mbox{co-dimension} three non-flat fibres give rise to
  higher order couplings in the effective four-dimensional
  superpotential---more specifically, in our example we find
  $\mathbf{10}\, \mathbf{5}\, \mathbf{5}\, \mathbf{5}$ couplings.
\end{abstract}

\thispagestyle{empty}

\clearpage


\newpage

\tableofcontents

\section{Introduction}

F-theory \cite{Vafa:1996xn} models have been extensively studied in
the last few years, starting with
\cite{Donagi:2008ca, Beasley:2008dc, Hayashi:2008ba, Beasley:2008kw, Donagi:2008kj},
for their promising features for GUT-inspired string theory model
building.

A detailed analysis of such models reveals that they sometimes develop
``non-flat'' points: these are points on the base over which the
dimension of the fiber jumps and, therefore, the standard
M-/F-theory\cite{Vafa:1996xn, Witten:1996qb} is not directly
applicable.\footnote{Examples of the appearance of such points in the literature go back to the early days of F-theory \cite{Candelas:2000nc} and showed up again with the advent of the intense study of non-abelian gauge groups together with $U(1)$ selection rules \cite{Braun:2013yti, Borchmann:2013hta}. For instance, two of the many examples appeared in the context of $SU(5)$-top constructions
over $\mathbb{P}_{[1,1,2]}$ \cite{Bouchard:2003bu, Braun:2013nqa,Borchmann:2013hta}.} In most phenomenological F-theory models,
where such loci would appear in the generic setting, they are excluded
from consideration by restricting the analysis to a highly non-generic
setup.

The goal of this note is to address the physical implications of such
non-flat point if they appear \emph{at co-dimension three}
\cite{Candelas:2000nc}. We will not give a general solution, but
rather analyze in detail a particular example with interesting
phenomenological properties. All the explicit details that we work out
in this article are obtained for the $SU(5)\times U(1)$ Peccei-Quinn
model which was already studied in \cite{Marsano:2009wr, Dolan:2011iu,
  Mayrhofer:2012zy}, and follow-up works, and which we review in
detail below. However, we expect related models to be amenable to an
analysis akin to the one we perform here.

Our main result is that in the weak coupling limit these non-flat
points do not interfere with the desirable GUT-physics, so they are
harmless for model building purposes. Indeed, the non-flat points
occur at special (self-)intersection points of the matter curves. As we show, they lead to higher order coupling built from the matter
states related to the (self-)intersecting curves. In our example at
hand, the $\ten_{3}$ curve meets the triple self-intersection of the
$\five_{-1}$ curve such that we observe a
$\ten_3 \five_{-1} \five_{-1} \five_{-1}$-coupling. The presence of
this coupling will not spoil the physics in a successful $SU(5)$ GUT
model. In fact, they will have very little effect, since the modes
involves will typically be massive.\footnote{We would like to thank the referee for
emphasizing this point to us.}

%
Before going into the analysis of non-flat points in our example, we
will resolve a small technical issue regarding $\mathbb Q$-factorial
terminal singularities which was not fully elucidated in
\cite{Mayrhofer:2012zy}. These are singularities at which uncharged
matter localizes. We will remove them by switching on complex
structure deformations, as in \cite{Arras:2016evy}.  This way, we do
not have to be concerned about these co-dimensional two effects when
we ultimately focus on the main topic of interest in this article, the
non-flat torus-fibrations at co-dimension three. Those come naturally
about when we relax the constraints on the base space of the F-theory
fibration which were imposed in \cite{Marsano:2009wr,
  Mayrhofer:2012zy}.
%
We find that in the resolved F-Theory four-fold the dimension of the
fibre over this point increases, i.e.\ the fibration becomes
non-flat. We study this co-dimension three effect from various angles,
and find that in each case we can interpret them as the above
mentioned higher order coupling. 

Along the way, we determine all the fluxes which
are either induced by matter curves \cite{Bies2017a} and the non-flat fibre. We calculate
the second Chern class of the fourfold and look at its implications on
the flux quantisation. We give the fluxes which must be turned on to
satisfy the quantisation condition and show that this flux forbids
string states in four dimensions, coming from M5 branes wrapping the
non-flat fibre.

\medskip

We have organized this paper as follows: in
section~\ref{sec:geometric-setup} we review the most relevant
geometric aspects of the global $SU(5)\times U(1)_{PQ}$ as studied in
\cite{Mayrhofer:2012zy}. Then we study the $\mathbb Q$-factorial
terminal singularities which appear in this setting and discuss how to
introduce complex structure deformations so that these singularities
do not appear. Afterwards we carefully analyse this fibration over a
general base without constraints. In section \ref{sec:fluxes}, we list
all the fluxes coming from the Mordell-Weil group, the matter
surfaces, and the non-flat fibres, respectively, and relate them with
the quantisation condition and explain why it forbids strings in
four-dimensions. In section~\ref{sec:IIB-picture}, we take the weak
coupling limit of our setting and study the states and their coupling
in the IIB picture. As a check, in section~\ref{sec:mirror-picture} we
go to the mirror/IIB side to confirm also from this perspective that
the non-flat point gives rise to a higher order coupling.  Finally, we
present our conclusions in section~\ref{sec:conclusions}.

\medskip

\emph{Note added:} As we were preparing this paper
\cite{Dierigl:2018nlv} appeared, which analyzes in detail the physics
associated to various non-flat fibrations in codimension two.

\section{The geometric setup}\label{sec:geometric-setup}

In this section, we review and extend the analysis of the global F-theory
realisation of the $SU(5)\times U(1)$ Peccei-Quinn model \cite{Mayrhofer:2012zy}, cf.~\cite{Marsano:2009wr, Dolan:2011iu} for the local description along the GUT-divisor. As a first step, let us first recall the
geometric setup presented in section 5 of \cite{Mayrhofer:2012zy}. 

To obtain the abelian $U(1)$ symmetry, we have to start from an elliptic fibration with Mordell-Weil group $\mathbb Z$ \cite{Morrison:1996na}. The Weierstra\ss{} model realising this symmetry takes the form \cite{Morrison:2012ei}
\begin{equation}
 \begin{aligned}\label{eq:Weierstrass-U1}
y^2 &= x^3 + (C_1 C_3 - B^2 C_0 - \frac{1}{3} C_2^2)x z^4 + \\
& + (C_0 C_3^2 - \frac{1}{3} C_1 C_2 C_3 + \frac{2}{27} C_2^3 - \frac{2}{3} B^2 C_0 C_2 + \frac{1}{4} B^2 C_1^2) z^6\,,
 \end{aligned}
\end{equation}
with the Mordell-Weil generator, i.e.\ the second section (besides the zero-section), given by:
\begin{equation}
(x,y,z) = (C_3^2 - \frac{2}{3} B^2 C_2^2, -C_3^3 +B^2 C_2 C_3 - \frac{1}{2} B^4 C_1,B)\,.
\end{equation}
%
As described in detail in section 5 of \cite{Morrison:2012ei}, one can resolve the co-dimensional singularties of \eqref{eq:Weierstrass-U1} by mapping it into a $\textmd{Bl}_{[0,1,0]}\mathbb{P}_{[1,1,2]}$ fibration:
\begin{eqnarray}\label{eq:MorrisonPark-U1}
\begin{aligned}
B_2\, v^2\,w + s\,w^2 + B_1\,s\,w\,v\,u &+ B_0\,s^2\,w\,u^2 = C_3\, v^3 u +\\
&+ C_2\,s\,v^2\,u^2 + C_1\,s^2\,v\,u^3 + C_0\,s^3\,u^4,
\end{aligned}
\end{eqnarray}
where $[u,\,v,\,w]$ are the homogeneous coordinates of
$\mathbb{P}_{[1,1,2]}$ and $s$ is the coordinate related to the
blow-up of
$\textmd{Bl}_{[0,1,0]}\mathbb{P}_{[1,1,2]}$.\footnote{\label{foo:fibre-taxonomy}
  The toric variety $\mathbb{P}_{[1,1,2]}$ is directly related to the
  6th two-dimensional reflexive polygon, in the standard ordering for
  such things cf.~\cite{Bouchard:2003bu,Braun:2013nqa}. Therefore, in
  the F-theory literature, cf.~\cite{Klevers:2014bqa}, sometimes
  \eqref{eq:MorrisonPark-U1} is known as the ``F6-fibration'', a
  nomenclature we will also use for brevity.} To obtain in addition to
the abelian symmetry the $SU(5)$-GUT including the Peccei-Quinn
symmetry, meaning that that the Higgs up and down multiplets can carry
different charges, we have to fix the sections
$B_2,\, B_1,\, B_0,\, C_0,\,\ldots,\,C_3$\footnote{As described in
  \cite{Morrison:2012ei}, to obtain the coefficients of
  \eqref{eq:Weierstrass-U1} from the coefficients of
  \eqref{eq:MorrisonPark-U1}, we must do a coordinate shift in $w$ to
  get rid of the linear terms, cf.\
  section~\ref{sec:weak_coupling_limit}.} in the following
way\cite{Marsano:2009wr, Dolan:2011iu, Mayrhofer:2012zy}:
\begin{equation}
\begin{aligned}\label{eq:coefficient-restriction}
 B_0 &= - \omega\, d_3\, \alpha = \omega\, B_{0,1}\,,\\
 B_1 &= - c_2\,d_3 = B_{1,0}\,,\\
 B_2 &= \delta = B_{2,0}\,,\\
 C_0 &= \omega^3\,\alpha\,\gamma = \omega^3\,C_{0,3}\,,\\
 C_1 &= \omega^2\,(d_2\,\alpha + c_2\,\gamma) = \omega^2\,C_{1,2}\,,\\
 C_2 &= \omega\,c_2\,d_2 = \omega\,C_{2,1}\,,\\
 C_3 &= \omega\,\beta = \omega\,C_{3,1}\,.
\end{aligned}
\end{equation}
Here $\alpha$, $\beta$, $\gamma$, $\delta$, $d_2$ ,$d_3$, $c_2$ are sections of line bundles of appropriate degree over the base.\footnote{For base manifolds which are given in terms of toric varieties or embeddings therein, these sections are homogeneous polynomials of certain (multi-)degrees.} The $I_5$-singular locus, i.e.\ the GUT-divisor, is at $\omega = 0$ on the base, as can be readily seen from plugging \eqref{eq:coefficient-restriction} into \eqref{eq:Weierstrass-U1} and taking the discriminat. 

Though \eqref{eq:MorrisonPark-U1} together with \eqref{eq:coefficient-restriction} pose the singular setting of the F-theory model we are interested in, we still have to resolve it to obtain a detailed understanding (via the duality to M-theory) of the physics of this setup.
As it turns out \cite{Mayrhofer:2012zy}, \eqref{eq:MorrisonPark-U1} plus \eqref{eq:coefficient-restriction} does not allow for a resolution of the fibration in a 
purely torical way. But we can still resolve parts of the hypersurface
singularities torically, and only for the final resolution step we have to introduce a
complete intersection to represent the smooth Calabi-Yau $\hat Y_4$. The resolved
model is given by the following two hypersurface equations
\begin{align}
\text{HSE}_1\,:&\quad \lambda_1 \, e  - \lambda_2 \, s \, P_2 = 0  \label{eq:hse1}\,,\\
\text{HSE}_2\,:&\quad \lambda_2 \,  Q  - \lambda_1 \, u\, P_1 = 0 \label{eq:hse2}\,,
\end{align}
with the polynomials
\begin{align}
 Q & = e_1\,s\,w^2 - e_4^2\, e_0\, \beta\, v^3\, u + e_4\,\delta\, v^2\, w\,,\\
 P_1 & = e_4\,e_0\,d_2\,u\,v + d_3\,w + e_1\,e_4\,e_0^2\,\gamma\,s\, u^2\,,\\
 P_2 & = c_2\, v + e_0\,e_1\,\alpha\,s\, u\,.
\end{align}
The homogeneous coordinates $[\lambda_1,\,\lambda_2]$ parameterize the $\mathbb{P}^1$, which was added in the final (small) resolution step.
To be more precise and for later reference, the two hypersurfaces \eqref{eq:hse1} and \eqref{eq:hse2} are embedded into the ambient variety with the relations
\begin{equation}
\begin{array}{cccccccccc|c c}
 u & v & w & s & e_0 & e_1 & e & e_4 & \lambda_1 & \lambda_2 & \textmd{HSE}_1 & \textmd{HSE}_2\\\hline
 1 & 1 & 2 & 0 & 0 & 0 & 0 & 0 & 1 & 0 & 1 & 4 \\
 0 & 1 & 1 & 1 & 0 & 0 & 0 & 0 & 2 & 0 & 2 & 3\\
 -c_B & 0 & 0 & [\delta] & [\omega] & 0 & 0 & 0 & 2[\delta] + [\omega] + [\alpha] - c_B & 0 & 2[\delta] + [\omega] + [\alpha] - c_B & [\delta]\\
 0 & 0 & -1 & 0 & -1 & 1 & 0 & 0 & 0 & 0 & 0 & -1 \\
 0 & -1 & -2 & 0 & -1 & 0 & 1 & 0 & -2 & 0 & -1 & -4\\
 0 & -1 & -1 & 0 & -1 & 0 & 0 & 1 & -1 & 0 & -1 & -2\\
 0 & 0 & 0 & 0 & 0 & 0 & 0 & 0 & 1 & 1 & 1 & 1
\end{array}
\end{equation}
for the homogeneous coordinates and the  Stanley-Reisner ideal:
\begin{equation}\label{eq:SR-Ideal}
 \textmd{SR-I} = \{u\,w,\, u\,e,\, u\,e_4,\, v\,s,\, v\,e_1,\, w\,e_0,\,
 s\,e_0,\, e_0\,e,\, \lambda_1\,\lambda_2,\, s\,e,\, s\,e_4,\, w\,e_4\}\,.
\end{equation}
Here $[\cdot]$ means the `degree' of the respective section or polynomial and $c_B$ is the `degree' of the first Chern class of the base space.

As noted in \cite{Mayrhofer:2012zy} this complete intersection Calabi-Yau (CICY) still has singularities. A careful analysis of \eqref{eq:hse1} and \eqref{eq:hse2} yields that there is a remaining singularity at the base loci
\begin{equation}\label{eq:Q-fac-terminal-locus}
\alpha = \gamma = 0
\end{equation}
and fibre coordinates $w=v=\lambda_1=0$.  Indeed, if we assume for the
above fibration a two-dimensional base then the so-obtained Calabi-Yau
threefold will be $\mathbb Q$-factorial with terminal singularity
points. Such varieties have recently been studied from the F-theory
perspective in \cite{Arras:2016evy, Grassi2018}. There it has been
pointed out that such singularities can only be resolved in a
discrepant way. Furthermore, upon compactification uncharged
hypermultiplets localise at these singularities which are needed to
cancel the six-dimensional gravitational anomaly.  It is not too
difficult to show that also the fibration at hand has the right amount
of uncharged singlets to be anomaly-free. The reader interested in the
explicit calculation is pointed to appendix
\ref{sec:anomaly-cancellation}.

Although these singularities are present in the original setup as presented in \cite{Mayrhofer:2012zy}, we can smooth them away by switching on complex structure deformations \cite{Namikawa1995}. Since the locus \eqref{eq:Q-fac-terminal-locus} lies generically away from the GUT-divisor, these deformation do not interfere with the local geometry at $\omega=0$ and only 
alter things away from it. Explicitly, we have to include the higher order terms
\begin{equation}\label{eq:cs-deformations}
B_{0,2},\, B_{1,1},\,C_{0,4},\,C_{1,3},\,C_{2,2},  
\end{equation}
in \eqref{eq:coefficient-restriction} which will give rise to
\begin{equation}
u^2\,w\,s^2\,,u\,v\,w\,s\,,u^4\,s^3\,,u^3\,v\,s^2,\,u^2\,v^2, 
\end{equation}
terms in $Q$, respectively. The thus obtained smooth geometry is the one we will study throughout the rest of the article.

\begin{figure}[h]
\tikzset{every picture/.style={line width=0.75pt}}
\begin{center}
  \includegraphics{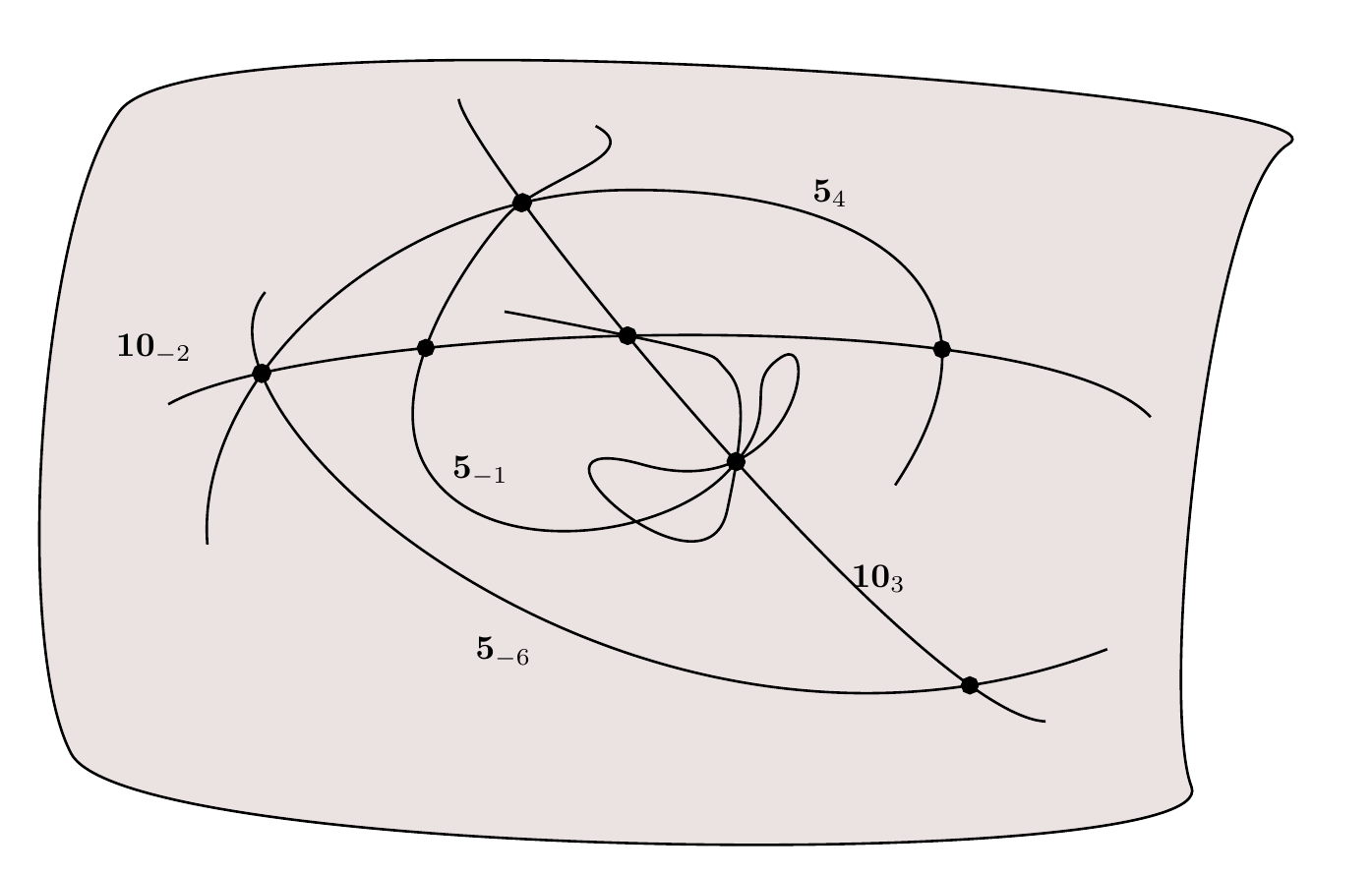}
\end{center}
 \caption{A sketch of the matter curves and Yukawa points within the $SU(5)$ GUT divisor $\{\omega = 0\}$. The seven bold dots indicate the six Yukawa points of \eqref{eq:non-abelian-Yukawa-pts} plus the triple intersection of the $\five_{-1}$-curve with the $\ten_3$-curve.}\label{fig:matter_and_Yukawas}
\end{figure}

Most of the details along the GUT-divisor of this $SU(5) \times U(1)_{PQ}$ fibration have been analysed in \cite{Mayrhofer:2012zy}. However, due to spectral cover considerations the locus
\begin{equation}\label{eq:non-flat-pts}
 \omega = \alpha = c_2 = 0 
\end{equation}
was excluded. But these loci are always presented if we consider the
above setting over a generic three-dimensional base. Therefore, we
examine these points very carefully in the following. However, we recall first
the most important features of the model.  We start with the two
$\ten$-curves:
\begin{equation}\label{eq:10-curves}
 \ten_{-2}:\quad d_3 = 0\,,\qquad
 \ten_3:\quad c_2 = 0\,,
\end{equation}
and the three $\mathbf{5}$-curves:
\begin{equation}
\begin{aligned}\label{eq:5-curves}
 \five_{-6}:&\quad \delta = 0\,,\\
 \five_{-1}:&\quad \alpha^2\,c_2\, d_2^2 + \alpha^3\,\beta\,d_3^2 + \alpha^3\,d_2\,d_3\,\delta - 2\,\alpha\,c_2^2\,d_2\,\gamma 
 - \alpha^2\,c_2\,d_3\,\delta\,\gamma + c_2^3\,\gamma^2 = 0\,,\\
 \five_4:&\quad \beta\, d_3 + d_2\,\delta = 0\,.
\end{aligned}
\end{equation}
The Yukawa-points at
\begin{equation}
 \begin{aligned}\label{eq:non-abelian-Yukawa-pts}
  \ten_{-2}\,\bar\five_{6}\,\bar\five_{-4}:& \quad \omega = d_3 = \delta = 0\,,\\
  \ten_{-2}\,\bar\five_1\,\bar\five_1: & \quad \omega = d_3 = \alpha\,d_2 - c_2\,\gamma\,,\\
  \ten_{3}\,\bar\five_{-4}\,\bar\five_{ 1}:& \quad \omega = c_2 = \beta\,d_3 + d_2\,\delta\,,\\
  \ten_{-2}\,\ten_{-2}\,\five_{4}:& \quad \omega = d_3 = d_2 = 0\,,\\
  \ten_{-2}\,\ten_{3}\,\five_{-1}:& \quad \omega = d_3 = c_2 = 0\,,\\
  \ten_{3} \,\ten_{3}\,\five_{-6}:& \quad \omega = c_2 = \delta = 0\,,
 \end{aligned}
\end{equation}
and
\begin{equation}
 \begin{aligned}\label{eq:abelian-Yukawa-pts}
  \bar\ten_{-3}\,\ten_{-2}\,\singlet_{5}:& \quad \omega = d_3 = c_2 = 0\,,\\
  \bar\five_{-4}\,\five_{-6}\,\singlet_{10}:& \quad \omega = \delta = \beta = 0\,,\\
  \bar\five_{1}\,\five_{-6}\,\singlet_{5}:& \quad \omega = \delta = \alpha^2\,c_2\, d_2^2 + \alpha^3\,\beta\,d_3^2 - 2\,\alpha\,c_2^2\,d_2\,\gamma 
  + c_2^3\,\gamma^2 = 0\,,\\
  \bar\five_{-4}\,\five_{-1}\,\singlet_{5}:& \quad \omega = \beta\,d_3 + d_2\,\delta = \alpha^2\,c_2\, d_2^2 - 2\,\alpha\,c_2^2\,d_2\,\gamma 
 - \alpha^2\,c_2\,d_3\,\delta\,\gamma + c_2^3\,\gamma^2 = 0\,,
 \end{aligned}
\end{equation}
have been presented in \cite{Mayrhofer:2012zy}. Besides these couplings, there is the intersection \eqref{eq:non-flat-pts} 
between the $\ten_{3}$-curve and the $\five_{-1}$-curve for which we cannot write down any gauge invariant three-point interaction. Looking at the second equation in \eqref{eq:5-curves}, we observe that the $\five_{-1}$-curve intersects the $\ten_{3}$-curve at the points \eqref{eq:non-flat-pts} three times, i.e.\ near $\alpha = c_2 = 0$ the $\five_{-1}$-curve takes the form
\begin{equation}
 (\alpha - \rho_1\,c_2)(\alpha - \rho_2\,c_2)(\alpha - \rho_3\,c_2) = 0
\end{equation}
with $\rho_i$ some constants. This hints already at a four-point coupling  $\ten_{3} \five_{-1} \five_{-1} \five_{-1}$ but to get a better picture of what really happens at these points, we have to look at the full fourfold geometry, especially the fibre structure. As it turns out, these are points where the dimension of the resolved fibre jumps, i.e.\ the fibration described by \eqref{eq:hse1} and \eqref{eq:hse2} over a three-dimensional (or higher dimensional) base is non-flat.\footnote{This does not imply that the dimension of the fourfold changes nor that it is singular at these points.} The dimensionality jump is due to the vanishing of $P_2$ at $\alpha = c_2 = 0$. We `lose' one of the equations which define the fibral curve of $E_3$
\begin{equation}
 E_3:\quad  e = P_2 = \lambda_2 \, Q - \lambda_1 \, P_1 = 0\,.
\end{equation}
A summary of the curves and the coupling points of this setup is depict in Figure~\ref{fig:matter_and_Yukawas}

\subsection{Fibre geometry at the non-flat points}
\begin{figure}
  \begin{center}
    \includegraphics{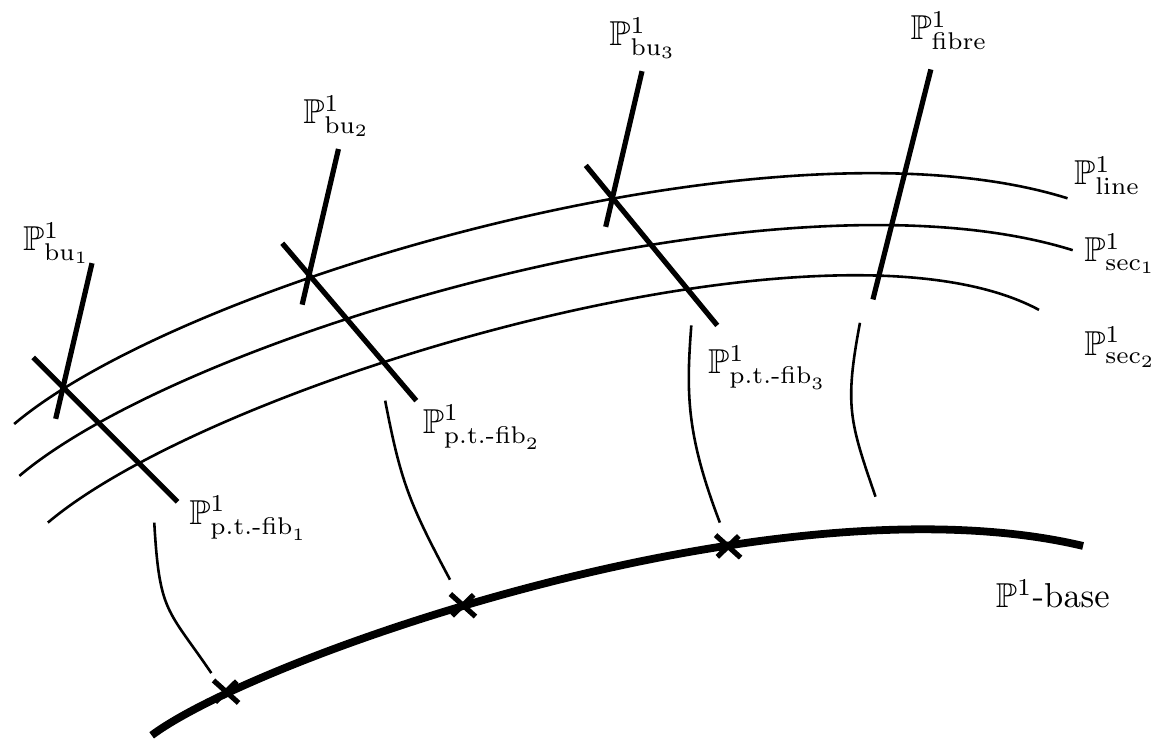}
  \end{center}
  \caption{Schematic drawing of the fibration structure of fibre surface $\FS$.}\label{fig:FS}
\end{figure}

Let us now present the details of the fibre above the non-flat points. At $\omega = \alpha = c_2 = 0$, the $\CP^1$-curves  of $E_1$, $E_3$ and $E_4$ split (or extend in dimension) in the following way
\begin{equation}
\begin{aligned}
 \CP^1_{E_1}\rightarrow \{&\Pnf{1}:\, e_1 = e_4 = \lambda_1 = 0,\,
 \Pnf{2}:\,e_1 = u\,e_0\,e_4\,\beta - w\,\delta = \lambda_1 = 0,\\
 & \Pnf{3}:\,e_1 = e = \lambda_2\,(u\,e_0\,e_4^2\,\beta - w\,e_4\,\delta) + \lambda_1\,(u^2\,e_0\,e_4\,d_2 + u\,w\,d_3) = 0\}\,,\\
 \CP^1_{E_3}\rightarrow \{& \FS:\, e = \lambda_2\,(v^3\,e_4^2\,\beta - v^2\,w\,e_4\,\delta - w^2\,e_1)
 + \lambda_1\,(e_1\,e_4\,\gamma + v\,e_4\,d_2 + w\,d_3) = 0\}\,,\\
 \CP^1_{E_4}\rightarrow \{& \Pnf{1}:\,e_1 = e_4 = \lambda_1,\,
 \Pnf{4}:\,e = e_4 = e_1\,\lambda_2 - \lambda_1\,d_3\}\,,
\end{aligned}
\end{equation}
whereas the fibres of $E_0$ and $E_2$ remain intact.\footnote{We should note here that for a different phase of the Coulomb branch, i.e.\ for another SR-ideal, the splitting can be different.} 
The Cartan charges of the above $\CP^1$'s are:
\begin{equation}
 \begin{aligned}
  \Pnf{1}:\quad & ( -1, 0, 1, -1)_{-3}\subset \bar\ten_{-3}\,,\\
  \Pnf{2}:\quad & ( -1, 0, 1, 0)_{3}\subset \ten_3\,,\\
  \Pnf{3}:\quad & ( 0, 1, -2, 1)_{0}\subset \text{roots}\,,\\
  \Pnf{4}:\quad & ( 1, 0, 0, -1)_{3}\subset \ten_3\,.
 \end{aligned}
\end{equation}
To see that the fibre surface $\FS$ at the non-flat points are del Pezzo four surfaces at a special complex structure sublocus, we give the reduced ambient space:
\begin{equation}\label{eq:ambient-space_at_non-flat-pt}
 \begin{array}{|c|c|c|c|c|c|c|c|}
 \hline
  v & e_1 & w & e_4 & \lambda_1 & \lambda_2 & \sum & \text{HSE}_2^{\text{red}}\\
  \hline
  1 & 1 & 1 & 0 & 2 & 0 & 5 & 3\\
  \hline
  0 & 0 & 1 & 1 & 1 & 0 & 3 & 2\\
  \hline
  0 & 0 & 0 & 0 & 1 & 1 & 2 & 1\\
  \hline
 \end{array}
\end{equation}
into which 
\begin{equation}\label{eq:hse2_red}
 \text{HSE}_2^\text{red}\,:\quad \lambda_2\,(v^3\,e_4^2\,\beta - v^2\,w\,e_4\,\delta - w^2\,e_1)
 + \lambda_1\,(e_1\,e_4\,\gamma + v\,e_4\,d_2 + w\,d_3)=0
\end{equation}
is embedded. The polynomials $\beta$, $\delta$, $\gamma$, $d_2$, $d_3$ from beforehand are now effectively coefficients. The toric space \eqref{eq:ambient-space_at_non-flat-pt} is a $\CP^1$-fibration over the Hirzebruch surface $\mathbb F_1\cong \text{dP}_1$ and \eqref{eq:hse2_red} defines a section of this fibration. Since the section degenerates over the points
\begin{equation}\label{eq:blow-up-pts}
 v^3\,e_4^2\,\beta - v^2\,w\,e_4\,\delta - w^2\,e_1 =
 e_1\,e_4\,\gamma + v\,e_4\,d_2 + w\,d_3 = 0\,,
\end{equation}
the del Pezzo one surface is blown up at three points. These three
points lie along a line. Therefore, the fibre-surface $\FS$ is not a
generic del Pezzo four surface but a degenerate dP$_4$. $\FS$ contains several rational curves: the generic fibre and the two special sections of $\mathbb F_1$; from the blow-ups of the Hirzebruch surface, we have the line going through the blown-up points, the exceptional $\CP^1$'s, and the proper transforms of the fibres at these points. The Cartan charges of the rational lines are:
\begin{equation}
 \begin{aligned}
  \CP^1_{\text{fibre}} & \cong \Pnf{3} \rightarrow ( 0,\, 1,\, -2,\, 1)_{0} \,,\\
  \CP^1_{\text{sec}_1} & = \{e = w = \text{HSE}_2^\text{red} = 0\}\rightarrow ( 1,\, 1,\, -2,\, 0)_{3} \,,\\
  \CP^1_{\text{sec}_2} & \cong \Pnf{4} \rightarrow ( 1,\, 0,\, 0,\, -1)_{3} \,,\\
  \CP^1_{\text{line}}  & = \{e = \lambda_2 = e_1\,e_4\,\gamma + v\,e_4\,d_2 + w\,d_3  = 0\}\rightarrow ( 1,\, -2,\, 1,\, 0)_{0} \,,\\
  \CP^1_{\text{bu}_i}  & =  \{e = 0 \wedge \eqref{eq:blow-up-pts}\}\rightarrow ( 0,\, 1,\, -1,\, 0)_{1} \,,\\
  \CP^1_{\text{p.t.-fib}_i} & \cong \CP^1_{\text{fibre}} - \CP^1_{\text{bu}_i} \rightarrow ( 0,\, 0,\, -1,\, 1)_{-1} \,.
 \end{aligned}
\end{equation}
Regarding $\CP^1_{\text{line}}$, we should note that prior to the blow-ups it was equivalent to $\CP^1_{\text{sec}_1}$, i.e.\ $\CP^1_{\text{line}}$ is the proper transform of sec$_1$ going through the three points which are blown-up. Hence, there are two special points for the complex structure deformation of $\CP^1_{\text{sec}_1}$; one where it splits into
\[
 \CP^1_{\text{sec}_1} \rightarrow \CP^1_{\text{line}} + \sum_{i=1}^3 \CP^1_{\text{bu}_i}
\]
and the one, which existed already in $\mathbb F_1$, where it becomes reducible to
\[
 \CP^1_{\text{sec}_1} \rightarrow \CP^1_{\text{sec}_2} + \CP^1_{\text{fibre}}\,.
\]
With these details at hand, we can describe the three-cycle which fuses three $\bar\five_{1}$ states into a $\ten_3$ state:
\begin{equation}
 \begin{aligned}
 (0,\, 1,\, -1,\, 0)_1& + (1,\, -1,\, 0,\, 0)_1  + (-1,\, 0,\, 0,\, 0)_1 \rightarrow
 \left(3\times (0,\, 1,\,-1,\, 0)_1 + (1,\,-2,\, 1,\, 0)_0\right) +\\
 &+(1,\,-2,\, 1,\, 0)_0 + (-2,\,1,\, 0,\, 0)_0 \rightarrow (1,\, 1,\, -2,\, 0)_3 +
 + (1,\,-2,\, 1,\, 0)_0 +\\
 &+ (-2,\,1,\, 0,\, 0)_0 \rightarrow (1,\, 0,\, 0,\, -1)_3 + (0,\,1,\, -2,\, 1)_0
 + (1,\,-2,\, 1,\, 0)_0 +\\
 &+ (-2,\,1,\, 0,\, 0)_0 \rightarrow (0,\, 0,\, -1,\, 0)_3\,.
 \end{aligned}
\end{equation}
In Figure~\ref{fig:FS}, we sketched $\FS$ to better
understand the interplay of the rational curves.

\section{Fluxes}\label{sec:fluxes}
Now that we have gained insight on the geometry of our  model, we can turn to the F-theory  four-form flux of our setup. It has to fulfil the flux quantisation
condition \cite{Witten:1996md, Freed:1999vc}:
\begin{equation} \label{eq:quantisation-condition}
 G_4 +\tfrac12 c_2(\hat Y_4)\in H^4(\hat Y_4,\mathbb{Z})\,,
\end{equation}
with $\hat Y_4$ the resolved Calabi-Yau four-fold. To see
whether \eqref{eq:quantisation-condition} forces us to switch on
half-integer fluxes, we are analysing  in the following the Chern class
of our four-fold \eqref{eq:hse1}--\eqref{eq:SR-Ideal}. The main goal of this study is to prove that
the restriction of $G_4$ to the non-flat fiber gives rise to a
non-trivial homology class. This fact provides a nice simplification
of the physics of the system, since it immediately implies that the M5
brane wrapping this divisor is inconsistent
\cite{Freed:1999vc}. Accordingly, the four dimensional light strings
this wrapped M5 would give rise to in four dimensions are
absent.\footnote{Note that even if the flux was trivial, this would
  not necessarily imply that the strings are light: they could still
  obtain a mass from periods of $C_3$, as conjectured in
  \cite{Collinucci:2016hgh} in a closely related case. But the
  existence of the flux makes the point moot.}

Let us also mention that non-trivial flux will potentially induce
chirality, and thus anomaly cancellation is a worry. Our goal in this
note is to clarify the dynamics arising from the non-flat
(codimension-three) point, while anomaly cancellation is a more global
phenomenon arising from matter curves, at codimension two. Therefore, we
expect our considerations to hold regardless of whether anomalies are
ultimately canceled in any specific model, as long as the local
behavior is as in our example.  Even
if not immediately relevant to us, the details of anomaly cancellation could be interesting. For example if there were underlying algebraic relations like the one observed in
\cite{Bies:2014sra}. We will leave such an analysis for
future work.

\subsection{The second Chern class}
Let us start by giving the second Chern class of an elliptically
fibred fourfold $\hat Y_4$ where the torus fibre is F6 (in the
nomenclature of footnote~\ref{foo:fibre-taxonomy}). For such
manifolds, where we did not impose an $SU(5)$ singularity yet, the
second Chern class reads\footnote{Here and in the following, we denote
  by capital letters the divisor class corresponding to the
  homogeneous coordinate given in terms of lower case letters,
  i.e. $U$ is the divisor class of the locus $\{u=0\}$. In the case of
  polynomials we use square brackets, i.e.\ $[c_2]$ denotes the
  divisor class with representative $\{c_2 = 0 \}$.}
\begin{equation}
 c_2(\hat Y_4) = \left(c_2(B_3) - c_1(B_3)^2\right) + 6\, c_1(B_3)\,(S + U + c_1(B_3) - [\delta])
 -[\delta]\,(S - U - c_1(B_3) - [\delta])\,.
\end{equation}
Hence, depending on the degree of $[\delta]$, when considering such an
F-theory compactification one might be forced to switch on flux even
though no non-abelian gauge groups are present yet. Note that this is
different from the $U(1)_X$ case \cite{Grimm:2010ez,Krause:2011xj}.

\subsection{\texorpdfstring{$U(1)$}{U(1)}- and matter surface fluxes}

Before coming to the second Chern class of the model including the
$SU(5)$, we write down the different fluxes which we can construct
from the Mordell-Weil generator and the matter surfaces. This helps us
in the next section to give the second Chern class in a concise way.

From the section $S$ we obtain via the Shioda map \cite{Shioda:1989, Wazir:2001, Park:2011ji} the following expression for the $U(1)$-flux:
\begin{equation}
 G_4^{U(1)}(\mathcal F) = \mathcal F\,(5\,(S - U - [\delta] - c_1(B_3)) 
 + 4\,E_1 + 3\,\Lambda_2 + 2\,(E - \Lambda_2) + E_4) 
\end{equation}
with $\mathcal{F}\in \pi^* H^{1,1}(B_3, \mathbb Z)$. To construct from the matter surfaces  gauge invariant fluxes, we follow a similar strategy to the one presented in \cite{Bies2017a}. That way we obtain the following fluxes:
\begin{equation}
\begin{aligned}
 G_4(\ten_{-2})  &= 5\,(E_1 - \Lambda_1)\,E_4 - (2\,c_1(B_3) - ([\delta] +[\alpha] + [\omega]))\times\\
                 &\qquad\qquad   \times(2\,E_1 - \Lambda_2 + (E - \Lambda_2) + 3\,E_4)\,,\\
 G_4(\ten_{3})   &= 5\,\Lambda_1\,E_4 - ([\delta] +[\alpha] + [\omega] - c_1(B_3))(3\,E_1 + \Lambda_2 - (E- \Lambda_2) +  2\,E_4)\,,\\
 G_4(\five_{-6}) &= 5\,E_1\,U - [\delta]\,(4\,E_1 + 3\,\Lambda_2 + 2\,(E - \Lambda_2) + E_4)\,,\\
 G_4(\five_{-1}) &= ([P_1] - \Lambda_2)([P_2] - \Lambda_1) + S\,[P_1] - (4\,c_1(B_3) -2\,[\delta] - 3\,[\omega] - [\alpha])\times\\
 & \qquad\qquad\times(-E_1 -2\,\Lambda_2 + 2\,(E - \Lambda_2) + E_4)\,,\\
 G_4(\five_{4})  &= 5\, (E_1\,([P_1] - \Lambda_2 - E_4) + \Lambda_1\,E_4) - (3\,c_1(B_3) - ([\alpha] + 2\,[\omega]))\times\\
                 &\qquad\qquad   \times(E_1 -3\,\Lambda_2 - 2\,(E -\Lambda_2) - E_4)\,.
\end{aligned}
\end{equation}

\subsection{Second Chern class of the \texorpdfstring{$SU(5)\times U(1)_{PQ}$}{SU(5)xU(1)-PQ} fourfold}
With all these expressions at hand, we can now finally give the second Chern class of the fourfold $\hat Y_4$ under consideration:
\begin{eqnarray}
 c_2(\hat Y_4)& = &\left(c_2(B_3) - c_1(B_3)^2\right) + 6\, c_1(B_3)\,(S + U + c_1(B_3) - [\delta]) +\nonumber\\& & \qquad\qquad - G_4^{U(1)}(\omega) - G_4(\ten_{-2}) - G_4(\five_{4}) - G_4^{\textmd{nf}} + \textmd{even terms} = \nonumber\\
         & = & G_4^{U(1)}(\omega) + G_4(\ten_{-2}) + G_4(\five_{4}) + G_4^{\textmd{nf}} + \textmd{even terms}\,.\label{eq:c2(Y4)SU5xU1}
\end{eqnarray}
Here $G_4^{\textmd{nf}}$ is the flux corresponding to the four cycle $\FS$:
\begin{equation}
 G_4^{\textmd{nf}} = [c_2]\,(E - \Lambda_2 - E_1) + E_1\,(\Lambda_1 - S)\,.
\end{equation}

The main properties of the $G_4^{\textmd{nf}}$ flux are that it does not break the $SU(5)$ gauge symmetry and it localises at the non-flat points. To see the second, we can integrate $G_4^\textmd{nf}$ over all algebraic two-cycles in $\hat Y_4$ which are accessible to us, i.e.
\begin{equation}
 \int_{\hat Y_4} G_4^\textmd{nf} \,\mathcal C_i = 0 
\end{equation}
with
\begin{equation}\label{eq:4-cycle-base}
 \mathcal C_i = \{\Gamma \, \tilde\Gamma,\,
 U\, \Gamma,\,
 S\, \Gamma,\,
 E_1\, \Gamma,\,
 \Lambda_2\,\Gamma,\,
 E\, \Gamma,\,
 E_4\,\Gamma,\,
 E_4\,\Lambda_2,\,
 U\, E_1\}_i,\qquad i=1,\ldots,9
\end{equation}
and
\begin{equation}
\int_{\hat Y_4} G_4^\textmd{nf} \mathcal C_{10} \ne 0\,,\qquad
\int_{\FS} G_4^\textmd{nf} = \int_{\hat Y_4} G_4^\textmd{nf}  \mathcal C_{11}\ne 0 
\end{equation}
where $\mathcal C_{10} = E_1\,E_4$ and $\mathcal C_{11}$ is the four-cycle of the non-flat fibre.
In equation \eqref{eq:4-cycle-base} $\Gamma$ and $\tilde\Gamma$ are place holders for all possible divisor classes pulled back from the base $B_3$.

Before we can make our main point of this section, we should first notice that all odd fluxes besides $G_4^\textmd{nf}$ appearing in \eqref{eq:c2(Y4)SU5xU1} do not localise at the non-flat points, i.e.
\begin{equation}
\int_{\FS} c_2(\hat Y_4)|_{\FS} = \int_{\FS} G_4^\textmd{nf}. 
\end{equation}
Therefore, we conclude that by \eqref{eq:quantisation-condition} there must be a non-trivial $G_4$ flux on $\hat Y_4$, whose restriction to $\FS$ cancels this contribution.

\section{The weak coupling limit and the IIB picture}\label{sec:IIB-picture}

As we will argue, the F-theory model of interest to us can be taken to
weak coupling without breaking any of the GUT symmetries, and without
encountering any special behavior along the way. Since we are
interested in computing a superpotential coupling, which is a
holomorphic quantity, we expect that the result of computing such
quantities at weak coupling remains valid all through moduli space.

\subsection{Weak coupling limit}\label{sec:weak_coupling_limit}
We recall from section~\ref{sec:geometric-setup} that the generic elliptic fibre with one free Mordell-Weil generater, i.e.
\begin{equation}
 c_0\,u^4 + c_1\,u^3\,v + c_2\,u^2\,v^2 + c_3\,u\,v^3 + b_0\,u^2\,w + b_1\,u\,v\,w + b_2\,v^2\,w + w^2 = 0\,,
\end{equation}
can be brought via a birational transformation into Tate form
\begin{equation}
y^2 + a_1 xyz + a_3 yz^3 = x^3 + a_2 x^2 z^2 + a_4 xz^4 + a_6 z^6,
\end{equation}
with
\begin{eqnarray}\label{eq:Tate-coefficients}
\nonumber
& a_1 &= b_1 \\
\nonumber
& a_2 &= - (b_2\, c_1 + b_0\, c_3) \\
& a_3 &= -(b_0\, b_2 + c_2) \\
\nonumber
& a_4 &=  (b_2^2\, c_0 + b_0\,b_2\,c_2 + c_1\,c_3) \\
\nonumber
&a_6& = -(b_2^2\,c_0\,c_2 - b_1\,b_2\,c_0\,c_3 + b_0\,b_2\,c_1\,c_3 + c_0\,c_3^2).
\end{eqnarray}
In analogy to \cite{Tate}, we define
\begin{eqnarray}\label{bsTate}
\nonumber
& \textbf{b}_2 & = a_1^2 + 4\, a_2 \\
& \textbf{b}_4 & = a_1 \, a_3 + 2\, a_2^2 \\
\nonumber
& \textbf{b}_6 & = a_3^2 + 4\, a_6.
\end{eqnarray}
To take the weak coupling limit, we proceed along the lines of Sen's original work \cite{Sen:1997gv} and require $\textbf{b}_2$, $\textbf{b}_4$, and $\textbf{b}_6$ to scale (at leading oder) like $\epsilon^0$, $\epsilon^1$, and $\epsilon^2$, respectively, as we take the limit $\epsilon\rightarrow 0$. One way to obtain that behaviour is to take
\begin{equation}\label{eq:our-wc-limit}
c_i \rightarrow \epsilon \, c_i,
\end{equation} 
in \eqref{bsTate}. Collecting the constant term in $\textbf{b}_2 = R +  O(\epsilon)$ the linear term in $\textbf{b}_4 = S \epsilon +  O(\epsilon^2)$ and the quadratic term in $\textbf{b}_6 = T \epsilon^2 +  O(\epsilon^3)$ we can write the discriminat in the weak coupling limit as
\begin{equation}
\Delta =  \frac{1}{4} R^2\,(-RT + S^2) \epsilon^2 + {O}(\epsilon^3) =:\epsilon^2\,R^2\,\Delta_{w.c.} + O(\epsilon^3),
\end{equation}
Plugging \eqref{eq:Tate-coefficients} into $\Delta_{w.c.}$, we obtain the rather lengthy polynomial 
\begin{eqnarray}
\nonumber
&\Delta_{w.c}  &\sim b_2\,( b_2^3\, c_0^2 - b_1\, b_2^2\, c_0\, c_1 + b_0\, b_2^2\, c_1^2 - 2 \,b_0\,b_2^2\, c_0\, c_2 + b_1^2\, b_2\, c_0\, c_2 + \\
&& - b_0\, b_1\, b_2\, c_1\, c_2 + b_0^2\, b_2\, c_2^2 + 3\, b_0\, b_1\, b_2\, c_0\,c_3 - 2 \,b_0^2\,b_2\,c_1\,c_3 - b_1^3\,c_0\,c_3 + \\\nonumber
&& + b_0\,b_1^2\,c_1\,c_3 - b_0^2\,b_1\,c_2\,c_3 + b_0^3\,c_3^2)\,.
\end{eqnarray}
This is the IIB D-brane locus (without the orientifold plane) for the generic F6-fibration if we take the weak coupling limit as in \eqref{eq:our-wc-limit}. The corresponding Calabi-Yau threefold is given 
by following double cover of $B_3$:
\begin{equation}
\xi^2 - R = 0\,,
\end{equation}
where the vanishing set $\{R=0\}$ defines the orientifold plane and the orientifold action is naturally induced by
\begin{equation}\label{eq:orientifold-invo}
\xi \longleftrightarrow -\xi\,.
\end{equation}
Now we restrict the section $b_i$ and $c_i$ to the case we are interested in, i.e.\ $SU(5)\times U(1)_{PQ}$ \cite{Mayrhofer:2012zy}:
\begin{equation}
 \begin{aligned}
    b_0 &= - \omega\,d_3\,\alpha  + b_{0,2}\,\omega^2\,,\quad
  & b_1 &= -c_2\,d_3  + b_{1,1}\,\omega\,,\quad
  & b_2 &= \delta\,,\\
    c_0 &= - \omega^3\,\alpha\,\gamma\,,\quad
  & c_1 &= - \omega^2\,(d_2\,\alpha + c_2\,\gamma)\,,\quad
  & c_2 &= - \omega\,c_2\,d_2\,,\\
    \quad
  & & c_3 &= - \omega\,\beta\,.
  & &\quad
 \end{aligned}
\end{equation}
where, for convenience, we switch on only one complex structure deformation compared with \cite{Mayrhofer:2012zy}, cf.\ equation \eqref{eq:cs-deformations}. Thus, we obtain
\begin{equation}\label{eq:hse_weak_coupling}
 \xi^2 + c_2^2\,d_3^2 +
 \omega\,(b_{1,1}^2\,\omega - 4\,b_{0,2}\,\omega\,\delta + 4\,\alpha\,\delta\,d_3 - 2\,b_{1,1}\,c_2\,d_3)= 0
\end{equation}
for the hypersurface of the Calabi-Yau threefold. We do not show the rather lengthy expression for $\Delta_{w.c.}$ because it will turn out that in suitable coordinates the polynomial factorizes and the loci of the brane-image brane pair become evident. We work now close to the singular point\footnote{The hypersurface \eqref{eq:hse_weak_coupling} has obviously more singularities than the one at \eqref{eq:location_conifold_sing}. There, is for instance, a co-dimension two singularity along $\xi=\omega=d_3=0$. However, we will ignore this singularity because first of all we are only interested in the vicinity of \eqref{eq:location_conifold_sing} and secondly we could either resolve it or chose a fibration where $d_3$ is constant. All the other co-dimension three singularities can be treated like \cite{Collinucci:2016hgh}, cf.\ below. }
\begin{equation}\label{eq:location_conifold_sing}
\xi = \omega = c_2 = \alpha = 0,
\end{equation}
where we expect the higher order coupling to arise. In particular, we assume that all $d_3$ and $\delta$ are non-vanishing close to the points of interest. We define now
\begin{equation}
(u,\,w,\,\sigma) := ( c_2\,d_3,\, b_{1,1}^2\,\omega - 4\,b_{0,2}\,\omega\,\delta + 4\,\alpha\,\delta\,d_3 - 2\,b_{1,1}\,c_2\,d_3,\, \omega),
\end{equation}
such that the ordinary double point singularity, or conifold, takes the form
\begin{equation}
\xi^2 = u^2  + \sigma w \,.
\end{equation}
We can represent this confold also in a toric way by introducing the homogeneous coordinate $\alpha_i$, $\beta_i$ with $i=1,\,2$ and scaling relation:
\begin{equation}\label{conifold}
\begin{array}{c|c|c|c}
\alpha_1 & \alpha_2 & \beta_1 & \beta_2 \\
\hline
1 & 1 & -1 & -1
\end{array}
\end{equation}
where
\begin{equation}
|\alpha_1|^2 + |\alpha_2|^2 - |\beta_1|^2 - |\beta_2|^2 = 0.
\end{equation}
The affine coordinates from above are expressed in terms of homogeneous ones as
\begin{eqnarray}
(\xi,\,u,\,\sigma,\,w) = (\tfrac{1}{2}(\alpha_1 \beta_2 - \alpha_2 \beta_1),\,\tfrac{1}{2}(\alpha_1 \beta_2 + \alpha_2 \beta_1),\,-\alpha_1 \beta_1,\, \alpha_2 \beta_2).
\end{eqnarray}
Furthermore, the orientifold involution \eqref{eq:orientifold-invo} acts now via
\begin{equation}\label{eq:orientifold-invo-homo}
\alpha_i \longleftrightarrow \beta_i \,.
\end{equation}

Using these two coordinate changes, we can rewrite the D-brane locus close to the point of interest as follows:
\begin{multline}
\Delta_{w.c.} \sim  \alpha_1^5 \, \beta_1^5 \, \Big((-2\,b_{1,1}^2\,\delta^2\,\gamma + 8\,b_{0,2}\,\delta^3\,\gamma + b_{1,1}^3\,\delta\,d_2 - 4\,b_{0,2}\,b_{1,1}\,\delta^2\,d_2 - b_{1,1}^3\,\beta\,d_3)\,\alpha_1^3 +\\
+ (- 2\,b_{1,1}^2\,\delta\,d_2 + 8\,b_{0,2}\,\delta^2\,d_2 + 6\,b_{1,1}^2\,\beta\,d_3)\,\alpha_1^2\,\alpha_2 + (8\,\delta^2\,\gamma - 4\,b_{1,1}\,\delta\,d_2 - 12\,b_{1,1}\,\beta\,d_3)\,\alpha_1\,\alpha_2^2 + \\
+ (8\,\delta\,d_2 + 8\,\beta\,d_3)\,\alpha_2^3\Big) \,
\Big((-2\,b_{1,1}^2\,\delta^2\,\gamma + 8\,b_{0,2}\,\delta^3\,\gamma + b_{1,1}^3\,\delta\,d_2 - 4\,b_{0,2}\,b_{1,1}\,\delta^2\,d_2 - b_{1,1}^3\,\beta\,d_3)\,\beta_1^3 +\\
+ (- 2\,b_{1,1}^2\,\delta\,d_2 + 8\,b_{0,2}\,\delta^2\,d_2 + 6\,b_{1,1}^2\,\beta\,d_3)\,\beta_1^2\,\beta_2 + \\ + (8\,\delta^2\,\gamma - 4\,b_{1,1}\,\delta\,d_2 - 12\,b_{1,1}\,\beta\,d_3)\,\beta_1\,\beta_2^2 
+ (8\,\delta\,d_2 + 8\,\beta\,d_3)\,\beta_2^3\Big)\,.
\end{multline}
This makes it obvious that the flavor brane/image brane pair are respectively located at
\begin{eqnarray}
&P_1 =  \eta_0 \,\alpha_1^3 + \eta_1\,\alpha_1^2 \alpha_2 + \eta_2\,\alpha_1 \alpha_2^2 + \eta_3\, \alpha_2^3 &= 0\\
&P_2 = \eta_0\,\beta_1^3 + \eta_1\,\beta_1^2 \beta_2 + \eta_2\,\beta_1 \beta_2^2 + \eta_3\,\beta_2^3  &= 0.
\end{eqnarray}
whereas the GUT stack and image-stack are at $\alpha_1=0$ and $\beta_1=0$, respectively. Locally the $\eta_i$'s are invertible and we treat them as if they were non-zero complex numbers. Under this assumption, we can further factorise the flavour branes to
\begin{eqnarray}
&P_1 = \Pi_{i = 1}^3 (A^{i} \alpha_1 + B^{i} \alpha_2) \,,\\
&P_2 = \Pi_{i = 1}^3 (A^{i} \beta_1 + B^{i} \beta_2)\,.
\end{eqnarray}
This implies that close to the point of interest there are three incoming flavor branes $A^{i} \alpha_1 + B^{i} \alpha_2$ each with their respective mirror $A^{i} \beta_1 + B^{i} \beta_2$. 

\subsection{Ext groups and Quiver theory}

In order to construct the resulting gauge theory, we need to specify
all branes participating at the point of interest. Following
\cite{Collinucci:2016hgh}, we employ the method of non-commutative
crepant resolutions \cite{Bergh}. This entails describing branes as
elements of the derived category of quasi-coherent sheaves on say
$Y_+$. Open string states between these are expressed in terms of
morphisms between such objects, which in turn are elements of so
called Ext groups. (For a review of the relevant background material
aimed at physicists, see \cite{Aspinwall:2004jr}.) We will first
briefly review the general form of the construction for the conifold
in \S\ref{sec:Ext-review}, and will then apply this construction to
our non-flat point in \S\ref{sec:non-flat-Ext}.

\subsubsection{Non-commutative crepant resolution of the conifold}

\label{sec:Ext-review}

Consider again the singular conifold, described by
\begin{equation}
\text{Spec } \left( \mathbb{C}[\xi,u,w,\sigma]/ \langle \xi^2 - u^2 -\sigma w \rangle \right).
\end{equation}
This, as we saw above, is a toric variety
\begin{equation}
\begin{array}{c|c|c|c}
\alpha_1 & \alpha_2 & \beta_1 & \beta_2 \\
\hline
1 & 1 & -1 & -1
\end{array}\,.
\end{equation}
The conifold has two small crepant resolutions which correspond in toric language to different subdivisions of its fan. 
These are also toric varieties with homogeneous coordinates $\alpha_1,...,\beta_2$ subject to the constraint
\begin{equation}\label{cfdres}
|\alpha_1|^2 + |\alpha_2|^2 - |\beta_1|^2 - |\beta_2|^2 = t.
\end{equation}
The two small resolutions are distinguished by the sign of $t$, and we denote them as $Y_{\pm}$ respectively. 
Applying the orientifold involution \eqref{eq:orientifold-invo-homo} to \eqref{cfdres}, we see that $t \leftrightarrow - t$, that is to say the two resolutions $Y_{\pm}$ are exchanged. This means that the resolution mode corresponding to the $\mathbb{P}^1$ is projected out.

It is, however, possible to describe D-branes on the singular space directly using its non-commutative crepant resolution \cite{Bergh}. By this we mean a non-commutative ring $A$
\begin{eqnarray}
A = \text{End}(M \oplus R),
\end{eqnarray}
where $R = \mathbb{C}[\xi,u,w,\sigma]/\langle \xi^2 - u^2  - \sigma w \rangle$ and $M$ is
\begin{equation}
M = \text{coker} \left( \psi: R^2 \longrightarrow R^2 \right).
\end{equation}
Here the map $\psi$ is given by
\begin{equation}
\psi = 
\left(
\begin{array}{cc}
\xi + u & \sigma \\
w & \xi - u
\end{array}
\right).
\end{equation}
Notice that one could also take
\begin{equation}
M = \text{coker} \left( \phi: R^2 \longrightarrow R^2 \right),
\end{equation}
with 
\begin{equation}
\phi = 
\left(
\begin{array}{cc}
\xi - u & -\sigma \\
-w & \xi + u
\end{array}
\right).
\end{equation}
Observe that
\begin{eqnarray}
\phi \psi = \psi \phi = (\xi^2 - u^2 - \sigma w) 
\left(
\begin{array}{cc}
1 & 0 \\
0 & 1
\end{array}
\right).
\end{eqnarray}
We do not want to delve into the details but simply state that $A$ is derived equivalent to $Y_{\pm}$. More concretely there is a correspondence
\begin{equation}
D^{b}(\text{mod}(A)) \cong D^{b} (\text{QCoh}(Y_{\pm})),
\end{equation}
cf.\ Theorem 5.1 in \cite{Bergh}. As is well established \cite{Aspinwall:2004jr}, one can view objects of $D^{b} (\text{QCoh}(Y_{\pm}))$ as D-branes in the B-model and morphisms between them correspond to open strings states. 

Using the dictionary laid out in \cite{Collinucci:2016hgh}, we will map certain (complexes of) $A$-modules to D-branes of interest. In order describe these effectively note that
\begin{equation}
A = \text{End}(M \oplus R) = \text{End}(R,R) \oplus \text{End}(A,A) \oplus \text{End}(A,M) \oplus \text{End}(M,A).
\end{equation} 
As a quiver we can represent $A$ as
\begin{equation}\label{Quivercfd}
\begin{tikzpicture}
\node[circle] (I0) at (-1,0) {$R$};
\node[circle] (I1) at (1,0) {$M$};	

\draw [->] (I0) edge[loop left, "$e_0$"] (I0);
\draw [->] (I1) edge[loop right, "$e_1$"] (I1);

\draw [->>] (I0) edge[bend left, "$\alpha_{1,2}$"] (I1);
\draw [->>] (I1) edge[bend left, "$\beta_{1,2}$"] (I0);
\end{tikzpicture}
\end{equation}
Here 
\begin{eqnarray}
& \text{End}(R,R) = \langle e_0 \rangle \cong R \\
& \text{End}(R,R) = \langle e_1 \rangle \cong R \\
& \text{End}(R,M) = \langle \alpha_1,\alpha_2 \rangle \\
& \text{End}(M,R) = \langle \beta_1, \beta_2\rangle,
\end{eqnarray}
as $R$-vector spaces. In particular, $e_i$ are idempotents. \\
Any module of $A$ can be encoded as a quiver representation. As laid out in \cite{Collinucci:2016hgh}, the basic representations from which one builds D7-branes are:
\begin{eqnarray}
& P_0 = e_0 A\,, \\
& P_1 = e_1 A.
\end{eqnarray}
These are linear combinations of paths ending at the left and right node of the quiver \eqref{Quivercfd}, respectively. Clearly morphisms from $P_0$ to $P_1$ are generated by $\alpha_{1,2}$ and from $P_1$ to $P_0$ by $\beta_{1,2}$. Together with the assignment
\begin{eqnarray}
&&P_0 \mapsto \mathcal{O} \\
&&P_1 \mapsto \mathcal{O}(1),
\end{eqnarray}
where $\mathcal{O}$ is the structure sheaf of the resolved conifold, we obtain for instance
\begin{equation}
\left( P_0 \stackrel{\alpha_1}{\longrightarrow} P_1 \right) 
\mapsto \left( \mathcal{O} \stackrel{\alpha_1}{\longrightarrow} \mathcal{O}(1) \right).
\end{equation}
Here the map $\alpha_1$ between the sheaves is nothing but the fiberwise multiplication by the homogeneous coordinate. The power of this approach is that computing Ext-groups between complexes of sheaves is easier in the setting of quiver representations. Since all relevant computations were already carried out in \cite{Collinucci:2016hgh}, we will not demonstrate them but only list the results in the following.

Fractional branes given by D1-branes wrapping the resolution divisor are given by
\begin{eqnarray}
&& S_0 = \mathbb{C} \langle e_0 \rangle \\
&& S_1 = \mathbb{C} \langle e_1 \rangle.
\end{eqnarray}
In terms of diagrams 
\begin{center}
\begin{tikzpicture}
\node[circle] (I0) at (-1,0) {$\mathbb{C}$};
\node[circle] (I1) at (1,0) {$\{0\}$};	

\draw [->] (I0) edge[loop left, "$e_0$"] (I0);
\draw [->] (I1) edge[loop right, "$0$"] (I1);

\draw [->>] (I0) edge[bend left, "$0$"] (I1);
\draw [->>] (I1) edge[bend left, "$0$"] (I0);
\end{tikzpicture}
\end{center}
and 
\begin{center}
\begin{tikzpicture}
\node[circle] (I0) at (-1,0) {$\{0\}$};
\node[circle] (I1) at (1,0) {$\mathbb{C}$};	

\draw [->] (I0) edge[loop left, "$0$"] (I0);
\draw [->] (I1) edge[loop right, "$e_1$"] (I1);

\draw [->>] (I0) edge[bend left, "$0$"] (I1);
\draw [->>] (I1) edge[bend left, "$0$"] (I0);
\end{tikzpicture}
\end{center}
We will later indicate the what the objects $S_0, S_1$ look like in $D^{b}(Y_{+})$. It is, however, convenient to define $I_0 = S_0[-1]$ and $I_1 = S_1[-1]$. Then we can represent the resolved conifold by
\begin{equation}
\begin{tikzpicture}
\node[circle] (I0) at (-1,0) {$I_0$};
\node[circle] (I1) at (1,0) {$I_1$};	

\draw [->] (I0) edge[loop left, "$e_0$"] (I0);
\draw [->] (I1) edge[loop right, "$e_1$"] (I1);

\draw [->>] (I0) edge[bend left, "$\alpha_i$"] (I1);
\draw [->>] (I1) edge[bend left, "$\beta_i$"] (I0);
\end{tikzpicture}
\end{equation}
This follows from the fact that the moduli space of representations of dimension $(1,1)$ is exactly the resolved conifold, see \cite{Collinucci:2016hgh} section 3.2.1. 

The brane/image brane pairs appearing in this paper are
\begin{eqnarray}
& F_0^i&= 
\begin{tikzcd}[column sep = huge]
\mathcal{O}  \arrow [r, "A^{i} \alpha_1 + B^{i} \alpha_2"]  & \mathcal{O}(1) 
\end{tikzcd} \in \text{Obj} \left(\mathcal{D}^b(Y_+) \right) \\
&F_1^i &=
\begin{tikzcd}[column sep = huge]
\mathcal{O}  \arrow [r, "A^{i} \beta_1 + B^{i} \beta_2"]  & \mathcal{O}(-1) 
\end{tikzcd} \in \text{Obj} \left(\mathcal{D}^b(Y_+) \right).
\end{eqnarray}
These correspond to D7 branes located at the \textbf{5} curve. To see this apply the cokernel to the relevant maps, which is commonly referred to as Tachyon condensation. Moreover there is one pair of objects corresponding to D7 branes located at the \textbf{10} curve
\begin{eqnarray}
&G_0 &= 
\begin{tikzcd}[column sep = huge]
\mathcal{O}  \arrow [r, "\alpha_1"]  & \mathcal{O}(1) 
\end{tikzcd} \in \text{Obj} \left(\mathcal{D}^b(Y_+) \right) \\
&G_1 &=
\begin{tikzcd}[column sep = huge]
\mathcal{O}  \arrow [r, "\beta_1"]  & \mathcal{O}(-1) 
\end{tikzcd} \in \text{Obj} \left(\mathcal{D}^b(Y_+) \right).
\end{eqnarray}
We also have fractional branes D(-1) instantons described by objects $I_0 = S_0[-1]$ and $I_1 = S_1[-1]$ where
\begin{eqnarray}
&S_0 &= 
\begin{tikzcd}[column sep = huge]
\mathcal{O}(2)  \arrow [r, " \left( \begin{array}{c} \beta_2 \\ -\beta_1 \end{array} \right)"]  & \mathcal{O}(1)^{\oplus 2} \arrow [r, "\left( \begin{array}{c} \beta_1 \text{,}  \ \beta_2 \end{array} \right)"] & \mathcal{O}
\end{tikzcd} \in \text{Obj} \left(\mathcal{D}^b(Y_+) \right) \\
&S_1 &= 
\begin{tikzcd}[column sep = huge]
\mathcal{O}(1)  \arrow [r, " \left( \begin{array}{c} -\beta_2 \\ \beta_1 \end{array} \right)"]  & \mathcal{O}^{\oplus 2} \arrow [r, "\left( \begin{array}{c} \beta_1 \text{,}  \ \beta_2 \end{array} \right)"] & \mathcal{O}(-1) \arrow[r] & 0
\end{tikzcd} 
\end{eqnarray}
For more details on this see Appendix A of \cite{Collinucci:2016hgh}.

We now study the open string states between these branes by computing certain Ext groups between elements of $\mathcal{D}^b(Y_{+})$, where $Y_+$ is one of the crepant small resolutions of the conifold. 
To this end consider the pair
\begin{equation}
F_0^i = 
\begin{tikzcd}[column sep = huge]
\mathcal{O}  \arrow [r, "A^i \alpha_1 + B^i \alpha_2"]  & \mathcal{O}(1) 
\end{tikzcd} \in \text{Obj} \left(\mathcal{D}^b(Y_+) \right).
\end{equation}
\begin{equation}
F_1^i =
\begin{tikzcd}[column sep = huge]
\mathcal{O}  \arrow [r, "A^i \beta_1 + B^i \beta_2"]  & \mathcal{O}(-1) 
\end{tikzcd} \in \text{Obj} \left(\mathcal{D}^b(Y_+) \right).
\end{equation}
The groups $\text{Ext}^i(F_0,F_1)$ were calculated in \cite{Collinucci:2016hgh}, but only for the value $(A,B) = (0,1)$.
We claim that these are isomorphic to our Ext groups as the two complexes
\begin{eqnarray}
&\begin{tikzcd}[column sep = huge]
\mathcal{O}  \arrow [r, "A \alpha_1 + B \alpha_2"]  & \mathcal{O}(1), 
\end{tikzcd} \\
&\begin{tikzcd}[column sep = huge]
\mathcal{O}  \arrow [r, "\alpha_2"]  & \mathcal{O}(1) 
\end{tikzcd},
\end{eqnarray}
are isomorphic in $\mathcal{D}^b(Y_+)$. To see this consider the following automorphism of the conifold 
\begin{equation}
f: (\alpha_1,\alpha_2,\beta_1,\beta_2) \mapsto (\alpha_1, \frac{1}{B}(\alpha_2 - A \alpha_1),\beta_1,\beta_2) \equiv (\tilde{\alpha}_1,\tilde{\alpha}_2,\beta_1,\beta_2).
\end{equation}
Observe that $A \tilde{\alpha}_1 + B \tilde{\alpha}_2 = \alpha_2$.
One readily checks that
\begin{equation}
\begin{tikzcd}[column sep = huge]
f^{*} \mathcal{O} \arrow[d, "\cong"] \arrow [r, "A \tilde{\alpha}_1 + B \tilde{\alpha}_2"]  & f^{*} \mathcal{O}(1) \arrow[d, "\cong"] \\
\mathcal{O}  \arrow [r, "\alpha_2"]  & \mathcal{O} (1).
\end{tikzcd}
\end{equation}
Similarly, we obtain an isomorphism
\begin{equation}
\begin{tikzcd}[column sep = huge]
g^{*} \mathcal{O}(-1) \arrow[d, "\cong"] \arrow [r, "A \tilde{\beta}_1 + B \tilde{\beta}_2"]  & g^{*} \mathcal{O} \arrow[d, "\cong"] \\
\mathcal{O}(-1)  \arrow [r, "\beta_2"]  & \mathcal{O},
\end{tikzcd}
\end{equation}
where
\begin{equation}
g: (\alpha_1,\alpha_2,\beta_1,\beta_2) \mapsto (\alpha_1, \alpha_2,\beta_1,\frac{1}{B}(\beta_2- A \beta_1)) \equiv (\alpha_1,\alpha_2,\tilde{\beta}_1,\tilde{\beta}_2).
\end{equation}
This implies that all Ext groups computed in \cite{Collinucci:2016hgh} are isomorphic to the ones we will need, e.g.
\begin{eqnarray}
& \text{Ext}^{j}(F_0,F_1) \cong (0,\mathbb{C}[\alpha_1 \beta_1],0,0) \,,\\
& \text{Ext}^{j}(F_1,F_0) \cong (0,\mathbb{C}[\beta_1 \alpha_1],0,0)\,.
\end{eqnarray}

\subsubsection{The non-flat point at weak coupling}
\label{sec:non-flat-Ext}

\begin{figure}
\centering
\begin{tikzpicture}
\node (I0) at (-1,0) {$I_0$};
\node (I1) at (1,0) {$I_1$};

\node (G1) at (-2,1)  {$G_1$};
\node (G0) at (-2,-1)  {$G_0$};
\node (F1) at (2,-1)  {$F_1^i$};
\node (F0) at (2,1)  {$F_0^i$};

\draw [->>] (I0) edge[bend left] (I1);
\draw [->>] (I1) edge[bend left] (I0);

\draw [dashed] (G0) edge[bend right = 20] (F1);
\draw [dashed] (F1) edge (F0);
\draw [dashed] (G1) edge[bend left = 20] (F0);
\draw [dashed] (G1) edge (G0);

\draw [->] (G1) edge[bend left] (I1);
\draw [->] (I0) edge (G1);

\draw [->] (G0) edge (I0);
\draw [->] (I1) edge[bend left] (G0);

\draw [->] (F0) edge[bend right] (I0);
\draw [->] (I1) edge (F0);

\draw [->] (I0) edge[bend right] (F1);
\draw [->] (F1) edge (I1);
\end{tikzpicture}
\caption{Quiver theory for GUT and flavor branes. Note that one should
  draw $F_0^1, F_0^2, F_0^3$ separately and connect to the other nodes
  as indicated. For the sake of clarity only one flavor brane/image
  brane is shown.}\label{fig:quiver}
\end{figure}
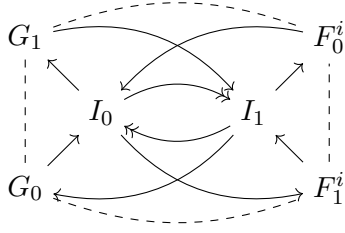

We now describe the relevant branes in our setup. There are the three pairs of objects
\begin{eqnarray}
& F_0^i&= 
\begin{tikzcd}[column sep = huge]
\mathcal{O}  \arrow [r, "A^{i} \alpha_1 + B^{i} \alpha_2"]  & \mathcal{O}(1) 
\end{tikzcd} \in \text{Obj} \left(\mathcal{D}^b(Y_+) \right) \\
&F_1^i &=
\begin{tikzcd}[column sep = huge]
\mathcal{O}  \arrow [r, "A^{i} \beta_1 + B^{i} \beta_2"]  & \mathcal{O}(-1) 
\end{tikzcd} \in \text{Obj} \left(\mathcal{D}^b(Y_+) \right).
\end{eqnarray}
These correspond to D7 branes located at the \textbf{5} curve. Moreover there is one pair of objects corresponding to D7 branes coming from the \textbf{10} curve
\begin{eqnarray}
&G_0 &= 
\begin{tikzcd}[column sep = huge]
\mathcal{O}  \arrow [r, "\alpha_1"]  & \mathcal{O}(1) 
\end{tikzcd} \in \text{Obj} \left(\mathcal{D}^b(Y_+) \right) \\
&G_1 &=
\begin{tikzcd}[column sep = huge]
\mathcal{O}  \arrow [r, "\beta_1"]  & \mathcal{O}(-1) 
\end{tikzcd} \in \text{Obj} \left(\mathcal{D}^b(Y_+) \right).
\end{eqnarray}
We also have fractional branes D(-1) instantons described by objects $I_0 = S_0[-1]$ and $I_1 = S_1[-1]$ where
\begin{eqnarray}
&S_0 &= 
\begin{tikzcd}[column sep = huge]
\mathcal{O}(2)  \arrow [r, " \left( \begin{array}{c} \beta_2 \\ -\beta_1 \end{array} \right)"]  & \mathcal{O}(1)^{\oplus 2} \arrow [r, "\left( \begin{array}{c} \beta_1 \text{,}  \ \beta_2 \end{array} \right)"] & \mathcal{O}
\end{tikzcd} \in \text{Obj} \left(\mathcal{D}^b(Y_+) \right) \\
&S_1 &= 
\begin{tikzcd}[column sep = huge]
\mathcal{O}(1)  \arrow [r, " \left( \begin{array}{c} -\beta_2 \\ \beta_1 \end{array} \right)"]  & \mathcal{O}^{\oplus 2} \arrow [r, "\left( \begin{array}{c} \beta_1 \text{,}  \ \beta_2 \end{array} \right)"] & \mathcal{O}(-1) \arrow[r] & 0
\end{tikzcd} 
\end{eqnarray}
A computation of the Ext groups shows \cite{Collinucci:2016hgh}
\begin{eqnarray}
& \text{Ext}^i(G_0,I_0) = (0,\mathbb{C},0,0), &\text{Ext}^i(G_0,I_1) = (0,0,\mathbb{C},0) \\
& \text{Ext}^i(G_1,I_0) = (0,0,\mathbb{C},0), &\text{Ext}^i(G_1,I_1) = (0,\mathbb{C},0,0) \\
& \text{Ext}^i(F_0,I_0) = (0,\mathbb{C},0,0), &\text{Ext}^i(F_0,I_1) = (0,0,\mathbb{C},0) \\
& \text{Ext}^i(F_1,I_0) = (0,0,\mathbb{C},0), &\text{Ext}^i(F_1,I_1) = (0,\mathbb{C},0,0),
\end{eqnarray}
and
\begin{eqnarray}
& \text{Ext}^i(G_0,G_1) \cong (0,\mathbb{C}[\alpha_2 \beta_2],0,0), &\text{Ext}^i(G_0,F_1) \cong (0,\mathbb{C}[\alpha_1 \beta_2],0,0) \\
& \text{Ext}^i(G_1,G_0) \cong (0,\mathbb{C}[\beta_2 \alpha_2],0,0), &\text{Ext}^i(G_1,F_0) \cong (0,\mathbb{C}[\beta_1 \alpha_2],0,0) \\
& \text{Ext}^i(F_0,F_1) \cong (0,\mathbb{C}[\alpha_1 \beta_1],0,0), &\text{Ext}^i(F_0,G_1) \cong (0,\mathbb{C}[\alpha_2 \beta_1],0,0) \\
& \text{Ext}^i(F_1,F_0) \cong (0,\mathbb{C}[\beta_1 \alpha_1],0,0), &\text{Ext}^i(F_1,G_0) \cong (0,\mathbb{C}[\beta_2 \alpha_1],0,0).
\end{eqnarray}
Also we have
\begin{equation}
\text{Ext}^1(I_1,I_0) \cong \text{Ext}^1(I_0,I_1) \cong \mathbb{C}^2.
\end{equation}
This situation is neatly summarized in a quiver diagram shown in  Figure~\ref{fig:quiver}.

In order to obtain the desired theory after orientifolding one takes the branes $G_i$ with multiplicity 5 to generate the GUT stack. A chiral bifundamental string between $G_0$ and $G_1$ giving rise to a state in the \textbf{10} representation upon orientifolding. This can be derived more rigorously by considering the gauge group on empty nodes. In \cite{Collinucci:2016hgh} it was shown that indeed we obtain $\Sp(0)$.

The flavor branes $F_j^i$ are each chosen with multiplicity 1. Between $G_1$ and each $F_0^j$ we have a bifundamental with the same chirality as above giving rise to a \textbf{5} state. Instanton effects arise from D1 branes wrapping the nodes $I_i$. We will only consider the case of a single instanton.

\begin{figure}
\centering
\begin{tikzpicture}
\node[circle] (ten) at (0,2) {$U(5)$};
\node[circle] (five1) at (-4,-2) {$U(1)$};
\node[circle] (five2) at (0,-2) {$U(1)$};
\node[circle] (five3) at (4,-2) {$U(1)$};

\node[circle, fill = yellow] (I0) at(-1.5,0) {$I_0$};
\node[circle, fill = yellow] (I1) at (1.5,0) {$I_1$};

\draw (ten) edge[bend right, "\textbf{5}$^1$"] (five1);
\draw (ten) edge["\textbf{5}$^2$"] (five2);
\draw (ten) edge[bend left, "\textbf{5}$^3$"] (five3);

\draw (ten) edge["$\lambda_1$"] (I1);
\draw (ten) edge[loop above,"\textbf{10}"] (ten);

\draw (five1) edge[bend right = 5,"$\nu_{11}$"] (I1);
\draw (five2) edge[bend right=30,"$\nu_{12}$"] (I1);
\draw (five3) edge["$\nu_{13}$"] (I1);

\draw [dashed] (five1) edge (I0);
\draw [dashed] (five2) edge (I0);
\draw [dashed] (five3) edge[bend left= 15] (I0);
\draw [dashed] (ten) edge (I0);
\draw [dashed] (I0) edge[bend right = 15] (I1); 
\draw [dashed] (I0) edge[bend left = 15] (I1); 
\end{tikzpicture}
\caption{Relevant zero modes for one D1 brane wrapping $I_1$ after orientifolding. Dashed lines indicate possible string states, but since $I_0$ is not occupied the play no relevance here. Labels such as $\nu_{12}$ refer only to bold lines. Note that we have orientifolded the quiver shown above.}\label{fig:ZeroModes}
\end{figure}
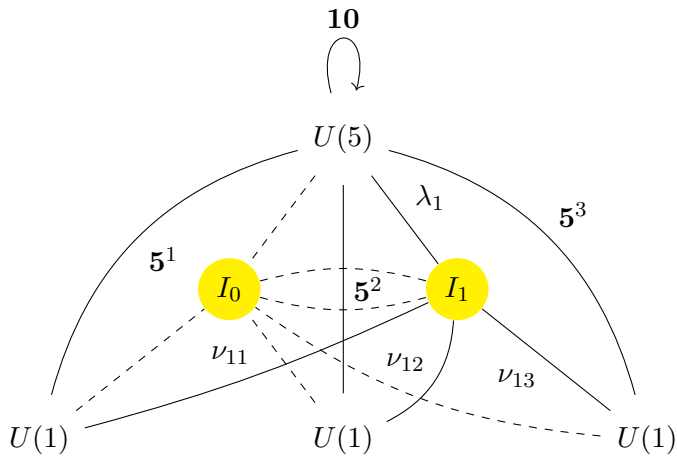

Firstly, consider a D1 brane wrapping $I_1$. This gives rise to charged zero modes as in Figure \ref{fig:ZeroModes}. Hence, the superpotential reads
\begin{equation}
W_{\text{inst}} = \lambda_1^i \textbf{10}^{\left[ij\right]} \lambda_1^j + \lambda_1^i \left( (\textbf{5}^1)^i \nu_{11} + (\textbf{5}^2)^i \nu_{12} + (\textbf{5}^3)^i \nu_{13} \right).
\end{equation}
Performing the integral 
\begin{equation}
\int d \lambda_1^i d \nu_{11} d \nu_{12} d \nu_{13} \exp(W_{\text{inst}}) = \textbf{10 5}^1 \textbf{5}^2 \textbf{5}^3,
\end{equation}
we obtain the desired coupling. If on the other hand we wrap one D1 brane around the $I_0$ node, there will be no contribution to the superpotential due to our choice of chirality.

\section{The mirror picture}\label{sec:mirror-picture}

Finally, it is interesting to see how the superpotential coupling
appears from the mirror IIA perspective. This mirror picture gives a
useful heuristic understanding of the physics, but the analysis is
harder to make fully precise than in the IIB setting, where we have a
well defined problem in algebraic geometry. The analysis is very
similar to that in \cite{Collinucci:2016hgh} (building on previous
work in \cite{Feng:2005gw,Forcella:2008au,Franco:2013ana}), so we will
be somewhat brief.

For the purposes of computing holomorphic data the topology of the
mirror to the conifold can be described by a fibration over $\bC$ with
fiber $\bC^*\times \Sigma$ \cite{Hori:2000kt,Hori:2000ck}, described
by
\begin{equation}
  \begin{split}
    uv & = W \, , \\
    P(x,y) & = W\, ,
  \end{split}
\end{equation}
where $W\in\bC$ parameterizes the base of the fibration, $u,v\in \bC$
parameterize the $\bC^*$ fiber, and $x,y\in \bC^*$ describe the
(punctured) Riemann surface $\Sigma$. For the specific case of the
conifold, we can choose a framing \cite{Bouchard:2011ya} such that
\begin{equation}
  P(x,y) = q + x + y + xy - xy^2\, .
\end{equation}
Here $q$ is a complex structure modulus mirror to the complexified
size of the small resolution of the conifold. This equation defines a
$\bP^1$ punctured at four points. As discussed in detail in
\cite{Feng:2005gw}, for the purposes of computing holomorphic quiver
data for our system, it is enough to focus our attention on $\Sigma$.

\begin{figure}
  \begin{center}
    \includegraphics[width=0.5\textwidth]{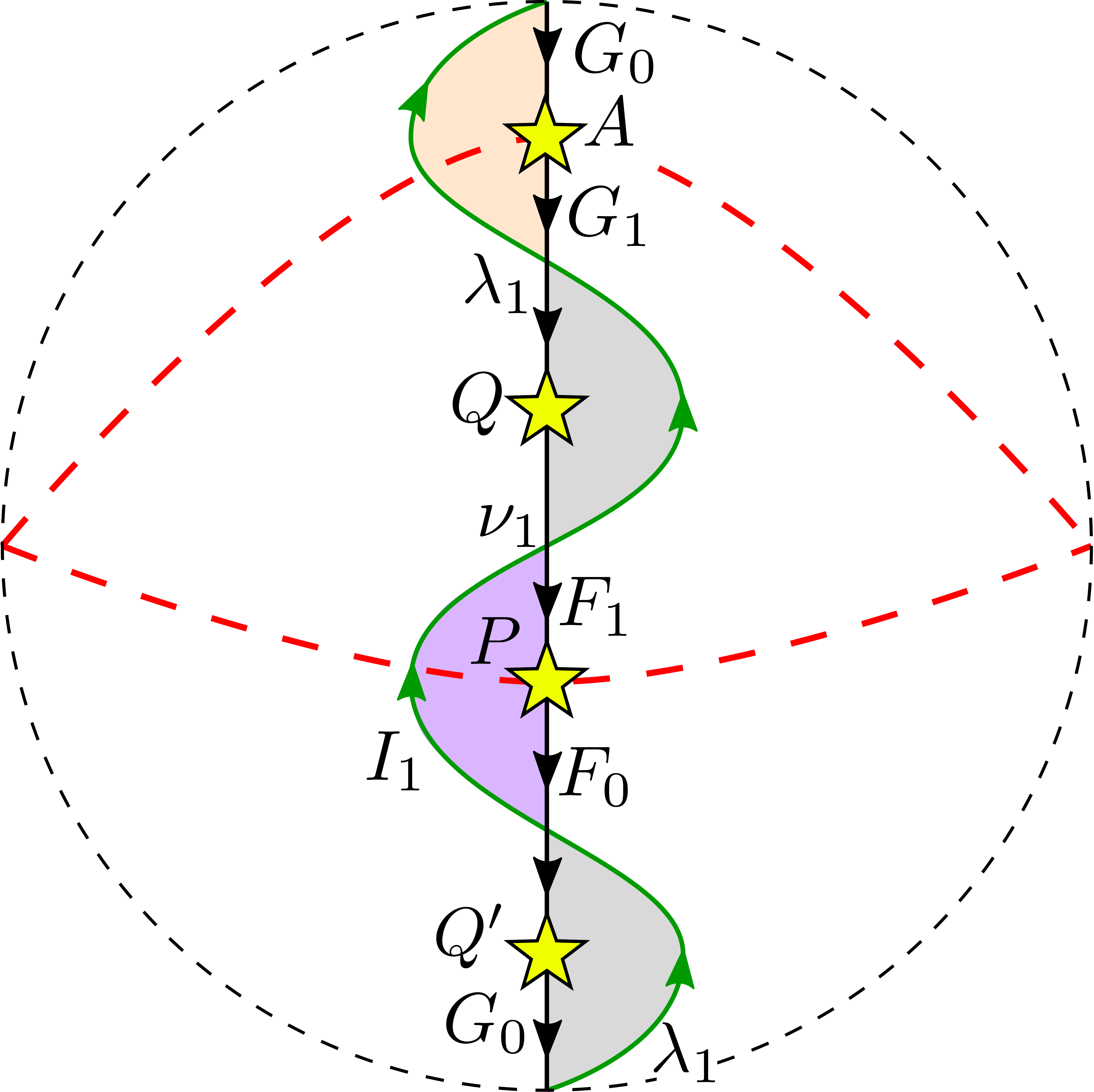}
  \end{center}
  \caption{Structure of branes and the orientifold involution on
    $\Sigma$. We outer dashed line should be identified with a point
    to obtain $\bP^1$. The four punctures have been marked by stars,
    and the orientifold involution induces a reflection along the red
    line (which becomes a reflection along the equator on $\bP^1$).}
  \label{fig:coniRiemann}
\end{figure}

In addition to the geometric background itself, we need to describe
how the branes wrap the geometry. The case with one $U(5)$ stack and
one additional $U(1)$ brane stack was described in detail in
\cite{Collinucci:2016hgh}. An important difference in our case is
that, in addition to the $U(5)$ stack, we have \emph{three} $U(1)$
flavor branes. We will start by analyzing the case in which all $U(1)$
branes are coincident, leading to a flavor stack with gauge group
$U(3)\times U(5)$. The restriction of the brane system to $\Sigma$ can
then be determined by identical arguments to those in
\cite{Collinucci:2016hgh}, with the result shown
in Figure~\ref{fig:coniRiemann}.

There are various features to note in Figure~\ref{fig:coniRiemann}. We
have the $G_0\sim G_1$ stacks (the identification is due to the
orientifold action), associated with the $U(5)$ stack, and the
$F_0\sim F_1$ stacks, associated to $U(3)$. We obtain various fields,
as these stacks intersect each other\footnote{Since the intersection
  is at a puncture, the existence of massless matter associated with
  the ``intersection'' is external input data from the point of view
  of the theory at the singularity.\label{footnote:global-matter}}, and additional matter fields as
the flavor stacks intersect the instanton brane $I_1$, with gauge
group $O(1)=\mathbb{Z}_2$. The resulting matter content can be summarized as
\begin{equation}
  \begin{array}{c|ccc}
    & U(5) & U(3) & O(1) \\
    \hline
    A & \mathbf{10} & \mathbf{1} & 0 \\
    Q & \mathbf{5} & \overline{\mathbf{3}} & 0 \\
    P & \mathbf{1} & \mathbf{3} & 0 \\
    \lambda & \overline{\mathbf{5}} & \mathbf{1} & 1 \\
    \nu & \mathbf{1} & \mathbf{3} & 1
  \end{array}\,.
\end{equation}
Note that $P$ is most naturally the (complex conjugate of the)
two-index representation of $SU(3)$, which can be identified with the
fundamental representation. The worldsheet instantons depicted in
Figure~\ref{fig:coniRiemann} then generate an effective action for the
charged instanton zero modes of the form
\begin{equation}
  S_{\text{inst}} = \lambda^i Q_{i}^{a} \nu_a + \lambda^i A_{[ij]}
  \lambda^j + \nu_a P^{[ab]} \nu_b
\end{equation}
where raising the index corresponds to going to the complex conjugate
representation. The effective non-perturbative superpotential one
obtains from integrating out the charged zero modes is then of the
form
\begin{equation}
  \label{eq:Wnp}
  W_{\text{np}} = \varepsilon_{abc}\varepsilon^{ijklm} A_{[ij]} Q_{k}^a
  Q_l^b Q_m^c + \varepsilon_{abc}\varepsilon^{ijklm} A_{[ij]}A_{[kl]}
  Q_m^a P^{[bc]} \cong AQ^3 + A^2PQ
\end{equation}
where we have omitted the unknown (but generically nonzero, since the
relevant worldsheet instantons have generically finite area)
coefficients of the various terms in the superpotential, which depend
on various geometric and brane moduli.

It is now a simple job to deform away from the $U(3)$ locus. This can
be seen as a Higgsing of the $SU(3)$ flavor symmetry, which will give
a mass to at least some of the fields in $P$, and generically to all
of them.\footnote{We will nevertheless keep the $Q$ fields
  massless. Recall from footnote~\ref{footnote:global-matter} that the
  massless spectrum of GUT fields is external data from the point of
  view of the singularity, which will be determined by global
  considerations.}  We can model this as the deformation
of~\eqref{eq:Wnp} given by
\begin{equation}
  W_{\text{np}} \to AQ^3 + A^2PQ + mP^2\, ,
\end{equation}
where, for simplicity, we have set all of the masses
equal. Integrating out $P$ then leads to an effective superpotential
of the form
\begin{equation}
  W_{\text{np}}' = AQ^3 - \frac{1}{4m} (QA^2)^2
\end{equation}
which in the $m\to\infty$ limit leads to the superpotential that we
have argued for in the previous section.

\section{Conclusions}\label{sec:conclusions}

The main focus of this note has been to understand non-flat fibres, in
co-dimension three, in F-theory and, in particular, their effect on
the low-energy dynamics. In previous analyses of the models that we
discuss here, this issue was sidestepped by drastically restricting
the base manifolds under consideration. Here we tied up this loose
end, by showing that it is not necessary to restrict the base
manifolds to avoid these points, as they are harmless for the good
phenomenological properties of the model.

Although we concentrated on one specific model for concreteness, it is
clear that the conclusions should hold fairly generally. This result
is significant for F-theory model building because non-flat points
seem to appear rather frequently. Hence, our result, that they are
harmless, does away with the need of having to worry about choosing
the base of the fibration with care in order to avoid such points, and
simplifies model building.

More explicitly, two of the many fibrations with non-flat points, at co-dimension three,
appeared in the context of $SU(5)$-top constructions over the fibre F6
\cite{Bouchard:2003bu, Braun:2013nqa, Borchmann:2013hta}, i.e.\ the third and fourth $SU(5)$-top \cite{Borchmann:2013hta} have a
non-flat points. 
Following our arguments, we see that in the case of the third $SU(5)$-top we
obtain the coupling $\ten_3\, \five_{-1}\five_{-1}\five_{-1}$ from
the non-flat points because the setup is almost identical to ours. For
the fourth $SU(5)$-top we expect the coupling
$\ten_{-1}\, \five_7 \,\five_{-3} \five_{-3}$.  A further example
shows up in the study of the exceptional gauge groups.  When looking
at the $E_6$-top over the Grimm-Weigand fibre \cite{Grimm:2010ez}, we
find again a non-flat fibre in co-dimension three. Though we do not
have a weak coupling limit in this case to carry out the second half
of our above analysis, we expect the appearance of a
$\mathbf{27}_{-1} \mathbf{27}_{-1} \mathbf{27}_{-1} \,\mathbf 1_3$
coupling. It would clearly be interesting to extend our analysis to
these models, and verify that indeed they do lead to harmless higher order couplings, as in the example that we have analyzed here.

\subsubsection*{Acknowledgements}

C.M.\ would like to thank Ling Lin, Eran Palti and Timo Weigand for
discussions and collaboration on a related project. 
The research of I.A.\ was supported by the IMPRS program of the MPP Munich
and  C.M.\ was funded in part by the Munich Excellence Cluster for Fundamental Physics ``Origin and the Structure of the Universe''.

\appendix

\section{6d anomaly cancellation}\label{sec:anomaly-cancellation}

We now verify the anomaly cancellation condition for the $\mathbb{Q}$-factorial Calabi-Yau threefolds with terminal singularities of section 2. As is very well explained in \cite{Arras:2016evy} a $\mathbb{Q}$-factorial variety $X$ is one where for each Weil divisor $D$ there exist an integer $n$ such that $nD$ is Cartier. If the resolved variety $\tilde{X}$ has canonical class $\tilde{K}$ and $X$ has canonical class $K$ then under the given circumstance
\begin{equation}
n \tilde{K} = f^{*}(n K) + n \sum a_i E_i,
\end{equation} 
for some integers $n, a_i$. Here $E_i$ are classes of the exceptional divisors. If $a_i >0$ for all $i$, then the singularities are called terminal.

Physically these singularities imply a localization of matter states from wrapped M2-branes, that is a number of uncharged hyper multiplets.

To verify the anomaly cancellation condition we have to compute the number of tensor, vector and hyper multiplets, $n_T,n_V,n_H$ arising from such a compactification. These have to satisfy
\begin{equation}
29 n_T - n_V + n_H = 273.
\end{equation}
We know that since we have an $SU(5) \times U(1)$ matter group
\begin{equation}
n_T = 0, \qquad n_V = 24 + 1 = 25.
\end{equation}
This leaves us with an unknown number of hyper multiplets $n_H$. As is well known these number splits up into number of uncharged $n_H^0$ and charged hyper multiplets $n_H^c$
\begin{equation}
n_H = n_H^0 + n_H^c.
\end{equation}
The charged hyper multiplets $n_H^c$ are counted by algebro-geometric means, and $n_H^0$ is computed via the topology of our variety.

\subsection{Counting charged hyper multiplets}

Charged hyper multiplets arise from so-called matter loci associated to gauge groups present in our theory. In the case at hand we have several charge $\ten$ and charge $\five$ loci as well as singlets.

We now restrict ourselves to working with
\begin{equation}
n = \text{deg}(\delta) = 2, \qquad \text{deg}(\alpha) = 1,
\end{equation}
which yields
\begin{eqnarray*}
&[\alpha]& \Rightarrow \text{deg}(\alpha)  = 1 \\
&[\beta] =  c_1(B_2) +  [\delta] - [\omega] & \Rightarrow \text{deg}(\beta) = 4 \\
&[\gamma] = 4 c_1(B_2) - 2[\delta] - 3 [\omega] -[\alpha] & \Rightarrow \text{deg}(\gamma)  = 4 \\
&[\delta] & \Rightarrow \text{deg}(\delta) = 2\\
& [c_2]   = [\delta] + [\alpha] + [\omega] -c_1(B_2) & \Rightarrow \text{deg}(c_2) = 1  \\ 
&[d_2] = 3 c_1(B_2) - [\delta] -2 [\omega] - [\alpha] & \Rightarrow \text{deg}(d_2)= 4  \\
&[d_3] = 2 c_1(B_2) - [\delta] -[\alpha] - [\omega] &\Rightarrow \text{deg}(d_3) = 2.
\end{eqnarray*}
Here we exploit the fact that over $B_2 = \mathbb{P}^2$ the degree of a homogenous ploynomial is equal to the first Chern class of its asssociated line bundle. The charge $\ten$ states are located at \eqref{eq:10-curves}.
It follows from Bezouts theorem that there are $\text{deg}(\omega)\cdot(\text{deg}(d_3) + \text{deg}(c_2)) = 3$ such points on the base. This gives us $30$ hyper multiplets.

The charge $\five$ loci are given by \eqref{eq:5-curves}. Applying Bezouts theorem again yields a total of $2 + 6 + 11 = 19$ such points. This gives a contribution of $5 \cdot 19 = 95$ multiplets.

Counting the number of singlets is more involved. We know \cite{Mayrhofer:2012zy} that the singlets of $U(1)$ charge  $\pm 10$ are located at
\begin{equation}
\delta = \omega \beta - \tfrac{1}{2} c_2 d_3 \delta = 0.
\end{equation}
This is equivalent to
\begin{equation}
\delta = \beta = 0.
\end{equation}
In our specific case Bezouts theorem implies that there are
\begin{equation}
\text{deg}(\delta) \cdot \text{deg}(\beta) = 2 \cdot 4 = 8,
\end{equation}
such points.

The singlets of $U(1)$ charge $\pm 5$ are located at the points satisfying
\begin{eqnarray}
\nonumber
F_1 :=&\beta  \text{c}_2^2 d_3^2 \delta ^2+\text{c}_2^2 d_2 d_3 \delta ^3-3 \beta ^2
\text{c}_2 d_3 \delta   \omega-2 \beta  \text{c}_2 d_2 \delta ^2  \omega\\
& +\gamma  \text{c}_2 \delta ^4  \omega+\alpha  \beta  d_3 \delta ^3  \omega+\alpha  d_2 \delta ^4  \omega+2 \beta ^3  \omega^2 =0,
\end{eqnarray}
\begin{eqnarray}
\nonumber
F_2 :=&-\alpha  \beta  \text{c}_2 d_3^2 \delta ^4-\alpha  \text{c}_2 d_2 d_3 \delta ^5+\beta
^2 \text{c}_2^2 d_3^2 \delta ^2+\\
&2 \beta  \text{c}_2^2 d_2 d_3 \delta^3+\text{c}_2^2 d_2^2 \delta ^4-2 \beta ^3 \text{c}_2 d_3 \delta   \omega\\
\nonumber
&-2 \beta ^2 \text{c}_2 d_2 \delta ^2  \omega+\alpha  \beta ^2 d_3 \delta ^3  \omega+\beta ^4  \omega^2 - \alpha \gamma  \delta ^6  \omega =0.
\end{eqnarray}
In addition points satisfying one of the following conditions must be excluded from this list:
\begin{eqnarray}\label{Conditions}
\nonumber
&\delta = \beta = 0\\
&\delta d_2 + \beta d_3 = 0\\
\nonumber
& c_2 = \omega = 0\\
\nonumber
& \delta = \omega = 0.
\end{eqnarray}
Generically the locus $F_1 = F_2 = 0$ consists of $14 \cdot 18$ points. We now subtract the points of \eqref{Conditions} weighted by their proper intersection multiplicity. This yields
\begin{equation}
14 \cdot 18 - 16 \cdot 2 \cdot 4 - 2 \cdot 6 - 1 \cdot 1 \cdot 1 - 10 \cdot 2 \cdot 1 = 91.
\end{equation}
All in all the number of uncharged hyper multiplets is
\begin{equation}
30 + 95 + 8 + 91 = 224.
\end{equation}

\subsection{Counting uncharged hyper multiplets}

The number of uncharged hyper multiplets is computed from the topological Euler characteristic and $h^{1,1}$ of our variety. We know that
\begin{equation}
h^{1,1} = 6.
\end{equation}
Strictly speaking this is the Hodge number of a smooth threefold rationally equivalent to our singular variety. The existence of such a deformation is guaranteed by \cite{Namikawa1995}.\\
The Euler characteristic of the singular variety is computed by first computing it for a smooth representative of its rational equivalence class. Then we use the fact that \cite{FultonInt}
\begin{equation}
\chi(X_{\text{Sing}}) -\chi(X_{\text{smooth}}) = \sum_P m_P,
\end{equation}
where the latter sum runs over the singular points $P$ and $m_P$ denotes the Milnor number of such a point. \\
The Euler characteristic  $\chi(X_{\text{smooth}})$ is computed using the toric embedding and turns out to be
\begin{equation}
\chi(X_{\text{smooth}}) = -132.
\end{equation}
We know that there is only one type of singularity located at
\begin{equation}
\alpha = \gamma = 0,
\end{equation}
which are $1 \cdot 4 = 4$ points. The Milnor numbers turn out to be 
\begin{equation}
m_P = 2.
\end{equation}
We thus end up with
\begin{equation}
\chi(X_{\text{Sing}}) =  \chi(X_{\text{smooth}}) + \sum_P m_P = -124.
\end{equation}
The number of uncharged multiplets then is simply
\begin{equation}
n_H^0 = 1 + h^{1,1} - \frac{1}{2} \chi(X_{\text{sing}}) + \frac{1}{2} \sum_P m_P = 7 + 62 + 4 = 73.
\end{equation}
We now add the universal hyper multiplet to that number to end up with
\begin{equation}
1 + n_H^0 = 74.
\end{equation}
We see that the anomaly cancellation condition is satisfied by computing
\begin{equation}
1 + n_H^0 + n_H^c - n_V = 1 + 73 + 224 -25 = 273.
\end{equation}

\bibliographystyle{utphys}
\bibliography{papers}

\end{document}